\documentclass[prd,superscriptaddress,twocolumn,showpacs,amsmath,
preprintnumbers,showkeys]{revtex4}
\usepackage{wrapfig,rotating}
\usepackage{graphicx,epsfig,color}

\makeatletter

\newenvironment{figurehere}
{\def\@captype{figure}}
{}
\makeatother

\usepackage{graphicx}
\usepackage{dcolumn}
\usepackage{bm}

\def\beq{\begin{equation}}
\def\eeq{\end{equation}}
\def\beeq{\begin{eqnarray}}
\def\eeeq{\end{eqnarray}}

\def\LQCD{\Lambda_{\mbox{\rm\scriptsize QCD}}}
\def\as{\alpha_{\mbox{\rm\scriptsize s}}}

\def\cO#1{{\cal{O}}\left(#1\right)}

\def\2GPD{$_2\mbox{GPD}$}
\def\GeV{{\rm Ge\!V}}
\def\oo{$1 \otimes 1$}
\def\12{$1\otimes 2$}
\def\22{$2 \otimes 2$}
\def\eff{{\mbox{\scriptsize eff}}}
\def\half{\frac{1}{2}}
\def\cD{{\cal{D}}}
\def\Qsep{Q_{\mbox{\rm\scriptsize sep}}}
\def\Qsep2{Q^2_{\mbox{\rm\scriptsize sep}}}

\begin{document}

\title{Perturbative QCD correlations in multi-parton collisions}

\author{B.\ Blok$^{1}$, Yu.\ Dokshitzer$^{2}$,   
L.\ Frankfurt$^{3}$
and M.\ Strikman$^{4}$
\\[2mm] \normalsize $^1$ Department of Physics, Technion -- Israel Institute of Technology,
Haifa, Israel
\\ \normalsize $^2$ CNRS, LPTHE, University Pierre et Marie Curie,
UMR 7589,  Paris, France
{\small On leave of absence: St.\ Petersburg Nuclear Physics Institute, Gatchina, Russia}\\
\normalsize $^3$ School of Physics and Astronomy,
\normalsize Tel Aviv University,
Tel Aviv, Israel
\\ \normalsize $^4$ Physics Department, Penn State University, University Park, PA, USA}

\begin{abstract}
We examine the role played in double parton interactions (DPI) by the parton--parton correlations originating from perturbative QCD parton splittings. 
Also presented are the results of the numerical analysis of the integrated DPI cross sections at Tevatron and LHC energies. 
To obtain the numerical results the knowledge of the single-parton GPDs gained by the HERA experiments was used
to construct the non-perturbative input for {\em generalized double parton distributions}. 
The perturbative two-parton correlations induced by three-parton interactions contribute significantly 
to resolution of the longstanding puzzle of an excess of multi-jet production events in the back-to-back kinematics observed at the Tevatron. 
\end{abstract}

  \maketitle
\thispagestyle{empty}

\vfill 

\section{Introduction}

{\em Multiple hard parton interactions}\/ (MPI) is an important element of the picture of strong interactions at high energies. 
The issue attracts a lot of attention. 
A series of theoretical studies were carried out in the last decade 
\cite{Treleani,Diehl,DiehlSchafer,Diehl2,Wiedemann,Frankfurt,Frankfurt1,SST,stirling,stirling1,Ryskin,Berger,BDFS1,BDFS2}. 
Attempts have been made to incorporate multi-parton collisions into event generators \cite{Pythia,Herwig,Lund}. 

Multi-parton interactions can serve as a probe for {\em non-perturbative correlations}\/ between partons in the nucleon 
wave function and are crucial for determining the structure of the underlying event at LHC energies.  
They constitute an important background for new physics searches at the LHC. 
A number of experimental studies were performed at the Tevatron \cite{Tevatron1,Tevatron2,Tevatron3}. 
New measurements are underway at the LHC \cite{Perugia,Fano}.

\smallskip
Double hard parton scattering in hadron--hadron collisions (DPI) can contribute to production
of {\em four}\/ hadron jets with large transverse momenta $p_\perp\gg \LQCD$, of {\em two}\/ electroweak bosons, 
or  ``mixed'' ensembles comprising three jets and $\gamma$, two jets and $W$, etc. 
In this paper we present the numerical results for a variety of final states.
However, for the sake of definiteness, in what follows we refer to production of four final state jets in the collision of    
partons 1 and 2 from one incident hadron with partons 3 and 4 from the second hadron: $1+3\to J_1 +J_3$, $2+4\to J_2 +J_4$.  
\medskip

Double hard interaction is a process difficult to approach theoretically. 
Formally speaking, it calls for analysis of four-parton operators that emerge in the squared matrix element describing a two-parton state in a hadron.
Relevant objects --- quasi-parton operators --- were introduced and classified, and evolution of their matrix elements 
studied by Bukhvostov et.\ al in the eighties in a series of papers \cite{Bukhvostov:1983te,Bukhvostov:1985vj,Bukhvostov:1987pr}.  

An approach to MPI based on the operator product expansion and on the notion of transverse momentum dependent 
parton distributions (TMD) is being developed in \cite{Diehl,DiehlSchafer, Diehl2}.   

MPI is a multi-scale problem. Not only because the separate parton--parton interactions may differ in hardness.
More importantly, {\em each}\/ single hard interaction possesses {\em two}\/ very different hardness scales. 
The distinctive feature of DPI is that it produces two pairs of nearly back-to-back jets, so that 
\begin{subequations}\label{eq:deltas}
\beeq
\delta_{13}^2 &\equiv& (\vec{J}_{1\perp} + \vec{J}_{3\perp})^2 \ll Q_1^2 = {J}_{1\perp}^2\simeq {J}_{3\perp}^2, \\
\delta_{24}^2 &\equiv& (\vec{J}_{2\perp} + \vec{J}_{4\perp})^2 \ll Q_2^2 = {J}_{2\perp}^2\simeq {J}_{4\perp}^2. 
\eeeq
\end{subequations}
Hence, in the collision of partons 1 and 3 the first (larger) scale is given by the invariant mass of the jet pair, 
$Q^2 = 4J_{1\perp}^2 \simeq 4J_{3\perp}^2$, while
the second scale is the magnitude of the {\em total transverse momentum}\/ of the pair: 
$\delta^2 = \delta_{13}^2$. 

\medskip

It is important to stress that in the MPI physics there is  {\em no factorization}\/ in the usual sense of the word. 
The cross sections do not factorize into the product of the hard parton interaction cross sections and the multi-parton distributions depending on momentum fractions $x_i$ and the hard scale(s). 

In \cite{BDFS1} we have introduced necessary theoretical tools for approaching the problem by introducing the {\em generalized double parton distributions}\/ ($_2\mbox{GPD}$s) in the momentum space. 
The double-parton GPD depends on one extra variable as compared to the double-parton distribution --- the transverse momentum mismatch $\vec\Delta$ between the partons in the wave function and the wave function conjugated.
In the mixed space of longitudinal momenta and transverse coordinates, an object equivalent to $_2\mbox{GPD}$ has been introduced by Treleani and Paver in the early eighties \cite{TreleaniPaver82}, and was long present in the literature since, see, in particular, \cite{TreleaniPaver85,mufti,Diehl}. 

In order to construct a viable model for the two-parton distribution stemming from the non-perturbative parton wave function of the proton and, in particular, for the $\Delta$-dependence of the corresponding part of $_2\mbox{GPD}$, we have exploited the existing experimental information obtained by the HERA experiments.  
This model disregards (rather arbitrarily, for lack of anything better) any {\em long-range correlations}\/ between the partons in the proton wave function, which approximation seems reasonable in a limited range of parton momenta  $10^{-1} >  x_i > 10^{-3}$ 
(see Section~\ref{SEC:Model} for details). 
This allows one to expresses the non-perturbative $_2\mbox{GPD}$ via the standard generalized parton distributions (GPDs) studied at HERA. 

\medskip
In \cite{BDFS2} we have studied how do perturbative QCD (pQCD) phenomena affect two-parton correlations. 
One obvious effect is ``scaling violation'' due to parton multiplication processes in higher orders.  
Another important effect is {\em short-range correlations}\/ induced by perturbative parton splitting, giving rise to 
{\em three-parton collisions}. Such contributions we will label as \12\
(``$3 \to4$'' of \cite{BDFS1,BDFS2}, ``$1\mbox{v}2$'' of  \cite{Gaunt_1v2}). 
This contribution to double hard interactions emerges as a result of collision of two partons 
from one hadron with the two offspring of the perturbative splitting of a single parton taken from another hadron. 
Perturbative splitting of one parton into two as a contributor to double-parton distributions was being discussed
in the literature for a long time, see, e.g., \cite{dDGLAP,Snigirev03}. 
However, for a relevant object --- double-parton GPD --- it was embedded into the parton evolution picture only recently.  

The pQCD expressions for fully differential distributions were derived in the leading collinear approximation in \cite{BDFS2}.  
In the present paper we examine how various contributions to the differential distribution, being integrated over transverse momentum imbalances, give rise to the expression for the {\em integrated cross section}\/ in terms of the product of $_2\mbox{GPD}$s of colliding hadrons.  

The results of numerical studies of the integrated DPI cross sections based on the theory and the model for non-perturbative two-parton distributions in the proton developed earlier are reported. 
We demonstrate that the \12\ processes contribute significantly to the DPI cross section 
and, in particular, are capable of explaining the longstanding Tevatron puzzle \cite{Tevatron1,Tevatron2,Tevatron3}.  
Namely, that the DPI cross section extracted by the the CDF and D0 at  $x_i\sim 0.01$ turned out to be a factor of two larger than the expectation based on the approximation of independent partons in the proton with the transverse distance spread extracted from the HERA data \cite{Frankfurt}.  

This observation does not exclude the presence of genuinely non-perturbative correlations between partons in the proton. However, we find it interesting that the pQCD induced correlations alone could explain the scale of the enhancement found by the Tevatron experiments. 

We also discuss the pattern of the $x$- and $Q^2$-dependence of the effective interaction area (aka ``effective cross section'') induced by pQCD correlations at Tevatron and LHC energies. 

\medskip

The paper is organized as follows. 

\noindent
In Section~\ref{SEC:Hidden} we reflect upon intrinsic complexity of theoretical MPI analysis. 
In Section~\ref{SEC:GPD} we remind the reader the basics of our approach to generalized two-parton distributions, $_2\mbox{GPD}$, and discuss the approximations adopted for constructing the model for $_2\mbox{GPD}$ in terms of the standard GPDs.  

In Section~\ref{SEC:P2PCOR} we discuss importance of perturbative parton correlations and give a semi-quantitative estimate of their magnitude. 
Section~\ref{SEC:12} is devoted to three-parton interactions (\12\  subprocesses). 
Here we reexamine the role of the so-called  ``short split'' term in \12\ subprocesses which is concentrated in the kinematical region of nearly equal transverse momentum imbalances,
\[ 
\delta'^2=(\vec{\delta}_{13}+\vec{\delta}_{24})^2 \ll \delta_{13}^2\simeq \delta_{24}^2 ,
\] 
see \cite{BDFS2}. 
We show here that the ``short split'' contribution to the {\em integrated cross section}\/ is actually contained in $\sigma_3$ that one obtains integrating the simple DDT-like formula derived in the complimentary kinematical region $\delta_{13}^2\ll \delta_{24}^2$ ($\delta_{13}^2\gg \delta_{24}^2$). 
Thus we correct (and simplify) the expression for the integrated DPI cross section (original Eq.~(28) of \cite{BDFS2} was plagued by double counting).   

In Section~\ref{SEC:Model} we discuss the independent parton model 
for the non-perturbative part of two-parton correlations.   
The numerical results for Tevatron and LHC energies are presented in Section~\ref{SEC:Numerics}. 
We conclude and discuss the results and perspectives of DPI studies in 
Section~\ref{SEC:CONC}.

\section{Hidden reefs of DPI physics \label{SEC:Hidden}}

An important question is, whether MPI admit an intuitive probabilistic parton interpretation as do the classical single hard interaction processes.  

When one considers inclusive one-parton distribution in a hadron (pdf), the quantum state of the parton in the light-cone hadron wave function (w.f.) coincides with that in the wave function conjugated (w.f.c.). This lays down the foundation for the probabilistic QCD improved parton picture. 

In the case of the two-parton correlation the situation is different. 
Here only the overall quantum characteristics of the parton pair as a whole (its total energy-momentum, its spin and color states) should be identical in the w.f.\  and w.f.c. 
As a result, the momentum, spin and color state of a single parton in the pair do not necessarily match in the wave function and in the wave function conjugated, thus endangering the very probabilistic interpretation of the process under consideration.     

A general approach to double hard interactions has been developed in \cite{BDFS1}.  
It turned out that the {\em transverse momentum}\/ of the parton in the w.f.\ and that of its counterpart in the w.f.c.\ are indeed necessarily different, with their difference $\vec{\Delta}$ being conjugate to the relative transverse distance between the two partons in the hadron. 
This has led to introduction of the notion of the {\em generalized double parton distribution}\/, $_2\mbox{GPD}$,  
which depends on a new momentum parameter $\vec{\Delta}$. 
It is important to stress that the cross section of the double hard process does not factorize 
into the product of the hard cross section and parton distributions. Instead, it contains a convolution of the product of two $_2\mbox{GPD}$s over $d^2\vec{\Delta}$.  

\medskip

Non-diagonality in the longitudinal momentum fractions is there too.  
Representing incident partons as plain waves with definite momenta, one ignores the fact that the two partons originate from one and the same finite size hadron. Therefore, when one picks two partons with momenta $x_1$, $x_2$ from the hadron wave function, one has to integrate over $x_1-x_2$ at the level of the amplitude, which integration ensures that the longitudinal separation between the partons does not exceed the size of the parent hadron. 
When one considers independent hard interactions of two pairs of partons, $(x_1,x_3)$ and $(x_2,x_4)$, taken from colliding hadrons, 
integrals over $r=x_1-x_2$ and $r'=x_3-x_4$ do not manifest themselves, since all four longitudinal momentum fractions $x_i$ are uniquely determined by the kinematics of the two produced hard systems. 

Not so for  \12\  processes. 
In this case a flow of large momenta between the two hard vertices is possible, and the integration over $r'$ is instrumental in getting rid of a fake singularity of the scattering matrix element (a simple explanation of the origin of this unphysical singularity can be found in \cite{singular}, see also \cite{BDFS2}). 

As a result, a small offset between the parton longitudinal momenta in the w.f.\ and w.f.c.\ emerges, $\left|{x_3-x_3'}\right|\propto \delta^2/Q^2$. 
However, this mismatch turns out to be negligible in the dominant kinematical region where
the squared total transverse momentum $\delta^2$ of the jet pair is much smaller than the overall hardness $Q^2$ of the process:  
$\delta_{13}^2\equiv (\vec{p}_{\perp 1}+\vec{p}_{\perp 3})^2\ll Q_1^2\simeq 4p_{\perp 1}^2$. 

\medskip
As noticed by Gaunt \cite{Gaunt13}, the non-diagonality in the longitudinal momentum space is likely to get induced in \12\ processes by higher order QCD effects. This happens when incoming partons, in the w.f.\ and w.f.c., exchange (real or virtual) gluons with transverse momenta in the interval $\LQCD^2\ll k_\perp^2\ll \delta^2$. 
In \cite{Gaunt13,Gaunt_MPItalk} arguments were raised in favor of smallness of the crosstalk effects in the DPI cross section.  
Otherwise, this would have led to unwelcome complication of the problem: one would be forced to deal with an unknown function of {\em four}\/ longitudinal momentum fractions (three independent variables) in place of the two of $_2\mbox{GPD}(x_1,x_2)$. 
Hence for the time being we disregard this complication.  

\medskip

We do not dwell into potential non-diagonality in color and in spin variables either.  
One may argue that such non-diagonal configurations are likely to be suppressed as they can be related with form factors 
for proton transition between two states with different quantum numbers of the proton constituents.  
For example, swapping inside the proton the colors of two partons that sit at distances of the order of the nucleon radius, 
corresponds to excitation that can be visualized as adding an extra piece of color string whose energy, $\cO{1\,\GeV}$, would excite and destroy the proton.  

\bigskip

\section{Generalized two-parton distribution  \label{SEC:GPD}}

In \cite{BDFS1,BDFS2} we have developed a formalism to address the problem of multi-parton interactions. 
QFT description of double hard parton collisions calls for introduction of a new  object --- 
{\em the generalized two-parton distribution}, $_2$GPD. 
Defined in the momentum space, it characterizes two-parton correlations inside hadron~\cite{BDFS1}:
\[
 D _h(x_1,x_2,Q_1^2,Q_2^2;\vec\Delta).
\] 
Here the index $h$ refers to the hadron, $x_1$ and $x_2$ are the light-cone fractions of the parton momenta, and $Q_1^2,Q_2^2$ the corresponding hard scales.  
As has been mention above, the two-dimensional vector $\vec\Delta$ is Fourier conjugate to the relative distance between the partons $1$ and $2$ in the impact parameter plane.
The distribution obviously depends on the parton species; we suppress the corresponding indices for brevity. 

The double hard interaction cross section (and, in particular, that of production of two dijets)
can be expressed through the generalized two-parton distributions $_2$GPD.

$_2\mbox{GPD}$s enter the expressions for the differential distributions in the jet transverse momentum imbalances $\vec{\delta}_{ik}$ 
in the kinematical region \eqref{eq:deltas}, as well as for the ``total'' DPI cross section (integrated over $\vec{\delta}_{ik}$). 
In the latter case the hardness parameters of the $_2\mbox{GPD}$s are given by the jet transverse momenta $Q_i^2$, 
while in the differential distributions --- by the imbalances $\delta_{ik}^2$ themselves. 
The corresponding formulae derived in the leading collinear approximation of pQCD can be found in Ref.~\cite{BDFS2}.
\medskip

It is important to bear in mind that the DPI cross section {\em does not factorize}\/ into the product of the hard parton interaction cross sections and the two two-parton distributions depending on momentum fractions $x_i$ and the hard scales, $Q_1^2, Q_2^2$. 
Instead,
\begin{widetext}
\begin{subequations}\label{eq:XsS}
\beeq
\frac{d\sigma^{\mbox{\scriptsize DPI}}}{dt_{13}dt_{24}} &=& \frac{d\sigma}{dt_{13}} \frac{d\sigma}{dt_{24}}\times \frac{1}{\sigma_\eff}, \\
\frac1{\sigma_\eff} &\equiv& \frac{ 
\int \frac{d^2\vec{\Delta}}{(2\pi)^2}  \> D_{h_1}(x_1,x_2, Q_1^2,Q_2^2;\vec\Delta) D_{h_2}(x_3,x_4, Q_1^2,Q_2^2; -\vec\Delta)}
{D_{h_1}(x_1,Q_1^2)D_{h_1}(x_2,Q_2^2)D_{h_2}(x_3,Q_1^2) D_{h_2}(x_4,Q_2^2)}.
\eeeq
\end{subequations}
\end{widetext}
The {\em effective interaction area}\/ $\sigma_\eff$ is given by the convolution of  the $_2\mbox{GPD}$s of incident hadrons 
over the transverse momentum parameter $\vec{\Delta}$ normalized by the product of single-parton inclusive pdfs.

\medskip
For expression \eqref{eq:XsS} for the DPI cross section to make sense, the integral over $\Delta$ has to be convergent. This is well the case when the two partons are taken from the non-perturbative (NP) proton wave function. Indeed, a typical inter-parton distance in the proton is large, of the order of the hadron size $R$. 
Accordingly, one expects the corresponding correlator in the momentum space to be concentrated at a finite
NP scale $\Delta^2\sim R^{-2}$ and to fall fast at large $\Delta^2$ (exponentially or as a sufficiently high power of $\Delta$). 

However, there is another source of two-parton correlations. This is purely perturbative (PT) mechanism when the two partons emerge from {\em perturbative splitting}\/ of one parton taken from the hadron wave function. 
In this scenario the production of the parton pair is concentrated at much smaller distances. 
As a result, the corresponding contribution to $_2$GPD turns out to be practically independent on $\Delta^2$ 
in a broad range, up to the hard scale(s) characterizing the hard process under consideration   
($\Delta^2$ only affects the lower limit of transverse 
momentum integrals in the parton cascades, causing but a mild logarithmic dependence).  

Given essentially different dependence on $\Delta$, one has to treat the two contributions separately by casting the $_2$GPD as a sum of two terms:
\beq\label{eq:2terms} 
\begin{split}
D_h  (x_1,x_2, Q_1^2,Q_2^2;\vec\Delta) &=  
{}_{[2]}D_h(x_1,x_2, Q_1^2,Q_2^2;\vec\Delta) \cr
& + {}_{[1]}D_h(x_1,x_2, Q_1^2,Q_2^2;\vec\Delta) .
\end{split}
\eeq
Here subscripts ${}_{[2]}D$ and ${}_{[1]}D$ mark the first and the second mechanisms, correspondingly: 
two partons from the wave function versus one parton that perturbatively splits into two.

\section{Perturbative two-parton correlations\label{SEC:P2PCOR}}

In this Section we discuss the role of the PT parton correlations and show that,  
given a sufficiently large scale of hard interactions, they turn out to be as important as NP ones.

 \subsection{Estimate of the PT correlation.}
 
Let us chose a scale $\Qsep2$ that separates NP and PT physics to be sufficiently low, so that parton cascades due to evolution between $\Qsep2$  and $Q_i^2$  are  well developed. 
To get a feeling of relative importance of the PT correlation, as well as to understand its dependence on $x$ and the ratio of scales, $Q^2$ vs.\ $\Qsep2$, the following lowest order PT estimate can be used.  

Imagine that at the scale $\Qsep2$ the nucleon consisted of $n_q$ quarks and $n_g$ gluons (``valence partons'') with relatively large longitudinal momenta, 
so that triggered partons with $x_1,x_2\ll1$ resulted necessarily from PT evolution. 
In the first logarithmic order, $\as\log (Q^2/\Qsep2)\equiv \xi$, the inclusive spectrum can be represented as
\[
    D \propto (n_qC_F + n_gN_c)\xi,
\]
where we suppressed $x$-dependence as irrelevant. 
If both gluons originate from the same ``valence'' parton, then 
\begin{subequations}
\beq\label{eq:1D}
{}_{[1]}\!D \propto  \frac12N_c\xi\cdot D \>+\,  (n_qC_F^2 + n_gN_c^2)\xi^2,
\eeq
while independent sources give ${}_{[2]}\!D$:
\beq\label{eq:2D}
\begin{split}
& \big(n_q(n_q\!-\!\!1)C_F^2 + 2n_qn_gC_FN_c +n_g(n_g\!-\!\!1)N_c^2\big)\xi^2 \cr 
& = D^2 - \big(n_qC_F^2 + n_gN_c^2\big)\xi^2 .
\end{split}
\eeq
\end{subequations}
Recall that the $\Delta$-dependence is different in \eqref{eq:1D} and  \eqref{eq:2D}. However, at $\Delta\!=\!0$
the second terms cancel in the sum and we get for the correlator
\beq\label{eq:corrmodel}
  \frac{D(x_1,x_2;0)}{D(x_1) D(x_2)} -1 \>\simeq\> \frac{N_c}{2(n_qC_F+n_gN_c)}.
\eeq
The correlation is driven by the gluon cascade ---- the first term in \eqref{eq:1D} --- and is not small (being of the order of unity).  
It gets diluted when the number of independent ``valence sources'' at the scale $\Qsep2$ increases. 
This happens, obviously, when $x_i$ are taken smaller. On the other hand, for large $x_i\sim 0.1$ and increasing, the effective number of more energetic partons in the nucleon is about 2 and decreasing, so that the relative importance of the \12 processes grows.

We conclude that the relative size of PT correlations is of order one, provided $\xi=\cO{1}$.
\medskip

Moreover, the PT parton correlations cannot be disregarded without running a risk of violating general principles. This can be illustrated  by looking at the momentum sum rule for double parton distributions.

\subsection{Momentum sum rule}

An obvious momentum sum rule should be satisfied. 
Namely, that the integral over $dx_2$ with the weight $x_2$ (summing over all parton species) should produce
in the end $(1-x_1)$ times the inclusive one-parton distribution $D(x_1)$, that is, the total longitudinal momentum 
carried by all the partons but the triggered one:
\beq\label{eq:momsum}
\begin{split}
\sum_{i_2} & \int dx_2\, x_2 \, D^{(i_1,i_2)}_h(x_1,x_2, Q_1^2,Q_2^2; \Delta=0) \cr
& \>=\> (1-x_1) \cdot D_h^{i_1}(x_1, Q_1^2)
\end{split}
\eeq
(here we have explicitly restored the parton species indices $i_1$, $i_2$).
This sum rule, together with other ones concerning valence quantum numbers, has been discussed by Gaunt and Stirling in \cite{stirling1} as means for restricting the form of double parton distributions. 
Setting the $\Delta$ argument of $_2\mbox{GPD}$ to $\Delta\!=\!0$ corresponds to taking integral over the relative transverse distance between partons in the proton.  

Derivation of \eqref{eq:momsum} is carried out in Appendix~\ref{SEC:Momentum}. 
It explicitly demonstrates that the PT parton splitting enters on equal grounds with the contribution due to two partons taken both from the initial NP hadron wave function.

\section{ \12\ DPI process \label{SEC:12}}

Actually, the NP and PT contributions {\em do not}\/ enter the physical DPI cross section in arithmetic sum \eqref{eq:2terms}, driving one even farther from the familiar factorization picture based on universal (process independent) parton distributions.  
As explained in \cite{BDFS2},  a double hard interaction of two pairs of partons that {\em both}\/ originate from PT 
splitting of a single parton from each of the colliding hadrons, does not produce back-to-back dijets.  
In fact, such an eventuality corresponds to a one-loop correction to the usual $2\to4$ jet production process
and should not be looked upon as multi-parton interaction. 
The term ${}_{[1]}D_{h_1}\times {}_{[1]}D_{h_2}$ has to be excluded from the product $D_{h_1}\times D_{h_2}$, 
the conclusion we share with Gaunt and Stirling \cite{Gaunt_1v2}.

So, we are left with two sources of genuine two-parton interactions: 
four-parton collisions described by the product of (PT-evolved) $_2$GPDs of NP origin (\22), 
\begin{subequations}\label{eq:2212}
\beq\label{eq:22}
{}_{[2]}\!D_{h_1}(x_1,x_2, Q_1^2,Q_2^2;\vec\Delta)\,  { } _{[2]}\!D_{h_2}(x_3,x_4,  Q_1^2,Q_2^2; -\vec\Delta) , 
\eeq
and three-parton collisions due to an interplay between the NP two-parton correlation in one hadron
and the two partons emerging from a PT parton splitting in another hadron (\12\ ), described by the combination 
\beq\label{eq:12}
\begin{split}
& _{[2]}\!D_{h_1} (x_1,x_2, Q_1^2,Q_2^2; \vec\Delta)
\, _{[1]}\!D_{h_2} (x_3,x_4, Q_1^2,Q_2^2; -\vec\Delta)
  \cr
+ \> & {}_{[1]}\!D_{h_1} (x_1,x_2, Q_1^2,Q_2^2; \vec\Delta)
\, _{[2]}\!D_{h_2}(x_3,x_4, Q_1^2,Q_2^2;-\vec\Delta) .
\end{split}
\eeq
\end{subequations}
Given that ${ }_{[2]}\!D$ falls fast at large $\Delta$, a mild logarithmic $\Delta$-dependence of ${}_{[1]}D$ can be neglected in the product in \eqref{eq:12}.

\subsection{On separation of NP and PT parts of ${ }_2\mbox{GPD}$. Parameter $Q_0^2$.}

Separation of PT and NP contributions is a delicate issue. 
By definition of the {\em perturbative}\/ correlation function, ${}_{[1]}D$ vanishes when $Q_1^2$, $Q_2^2$ are taken equal to the separation scale $\Qsep2$ that one chooses to set the lower limit for applicability of pQCD calculations.  
Strictly speaking, $\Qsep2$ can be chosen arbitrarily: both the NP input function ${ }_{[2]}\!D$ and the PT-calculable correlation ${}_{[1]}D$ contain $\Qsep2$-dependence, but their sum 
does not depend on this formal parameter. 
At the same time, the character of the $\Delta$-dependence of ${ }_{[2]}\!D$ depends, obviously, on the choice of the $\Qsep2$ scale. Indeed, by increasing the value of $\Qsep2$ one will shuffle a part of the perturbative splitting contribution from ${ }_{[1]}\!D$ into ${ }_{[2]}\!D$. 
As a result, the ``NP correlator''  ${ }_{[2]}\!D$, contaminated by a short range PT correlation, would acquire a ``tail'' at large $\Delta^2$ which would spoil convergence of the $\Delta$ integration in \eqref{eq:XsS}. 

Thus, in order to preserve the logic of the NP--PT separation,  
one is led to introduce a specific resolution scale, $\Qsep2=Q_0^2$, 
at which scale the NP correlation ${ }_{[2]}\!D$ falls fast with increase of $\Delta^2$.  
So defined, $Q_0^2$ is no longer an arbitrary ``factorization scale'' but a {\em phenomenological parameter}\/ whose value (which one expects to be of order of $1 \GeV$) should be established from the data.

\subsection{Composition of the \12 DPI cross section}

In order to derive the DPI cross section, one has to start with examination of the double differential  transverse momentum distribution and then integrate it over jet imbalances $\delta_{ik}$. 
Why this step is necessary? 

The parton distribution  $D(x,Q^2)$ --- the core object of the QCD-modified parton model ---  arises upon logarithmic integration over the transverse momentum up to the hard scale, $k_\perp^2< Q^2$.
Analogously, the double parton distribution $D(x_1,x_2,Q_1^2,Q_2^2; \vec\Delta)$ embeds {\em independent integrations}\/ over parton transverse momenta $k_{1\perp}^2$, $k_{2\perp}^2$ up to $Q_1^2$ and $Q_2^2$, respectively. 
However, the \12\ DPI cross section contains a specific contribution (``short split'', see below) in which the transverse momenta of the partons 1 and 2 are strongly correlated (nearly opposite). This pattern does not fit into the structure of the pQCD evolution equation for ${ }_2\mbox{GPD}$ where $k_{1\perp}$ and $k_{2\perp}$ change independently. 
Given this subtlety, a legitimate question arises whether the expression for the integrated \12\ cross section \eqref{eq:12} based on the notion of the two-parton distribution ${ }_{[1]}\!D$ 
takes the short split into account. 
Below we demonstrate that in fact it does.   

\medskip 

The differential distribution over jet imbalances was derived in \cite{BDFS2} in the leading collinear approximation of pQCD. 
It resembles the ``DDT formula'' for the Drell-Yan spectrum \cite{DDT} and 
contains two derivatives of the product of $_2\mbox{GPD}$s \eqref{eq:2212} that depend on the corresponding $\delta_{ik}$ as hardness scales, and the proper Sudakov form factors depending on (the ratio of) the $Q_i^2$ and $\delta^2_{ik}$.  

In particular,  in the region of {\em strongly ordered}\/ imbalances,
\beq\label{eq:dlog1}
\frac{\pi^2 d\sigma^{\mbox{\scriptsize DPI}}}{d^2\delta_{13} \, d^2\delta_{24}} \propto \frac{\as^2}{\delta_{13}^2 \, \delta_{24}^2}; \quad
\delta_{13}^2 \>\gg\> \delta_{24}^2, \>\>   \delta_{13}^2 \>\ll\> \delta_{24}^2,  
\eeq
the differential \12\ cross section reads 
\begin{widetext}
\beeq\label{eq:DDT31}
\frac{\pi^2\, d\sigma_{\mbox{\scriptsize\12}}}{d^2\delta_{13}\, d^2\delta_{24}} &=&
 \frac {d\sigma_{{\mbox{\scriptsize part}}}
 } {d\hat{t}_1\,d\hat{t}_2}
\cdot \frac{d}{d\delta_{13}^2} \frac{d}{d\delta_{24}^2} \bigg\{   \int\!\! \frac{d^2\vec{\Delta}}{(2\pi)^2} \>
{}_{[1]}\!D_{h_1}(x_1,x_2,\delta_{13}^2,\delta_{24}^2; \vec{\Delta}) \> {}_{[2]}\!D_{h_2}(x_3,x_4,\delta_{13}^2, \delta_{24}^2; \vec{\Delta} )
\nonumber\\
&\times&  S_1\left({Q_1^2},\delta_{13}^2 \right)S_3\left({Q_1^2},{\delta_{13}^2}\right)
 \cdot S_2\left({Q_2^2},{\delta_{24}^2}\right) S_4\left({Q_2^2},\delta_{24}^2\right)   \bigg\} +  \big\{ h_1 \leftrightarrow h_2 \big\}.
\eeeq
\end{widetext}
The differential distribution for the \22\ DPI mechanism has a similar structure, see Eq.~(25) of \cite{BDFS2}. 

In addition to \eqref{eq:DDT31}, there is another source of double collinear enhancement in the differential \12\ cross section. 
It is due to the kinematical region where the two imbalances nearly compensate one another,
\beq\label{eq:deltaprime}
\delta'^2 = (\vec{\delta}_{13} + \vec{\delta}_{24})^2 \>\ll\> \delta^2 = {\delta}_{13}^2\simeq {\delta}_{24}^2,
\eeq
and the dominant integration region is complementary to that of \eqref{eq:dlog1}:
\beq\label{eq:dlog2}
\frac{\pi^2 d\sigma^{\mbox{\scriptsize DPI}}_{\mbox{\scriptsize short}}}{d^2\delta_{13} \, d^2\delta_{24}} \propto \frac{\as^2}{\delta'^2 \, \delta^2}; \qquad
\delta'^2 \>\ll\> \delta^2 .
\eeq

This enhancement characterizes the set of \12\  graphs in which there is no accompanying radiation 
with transverse momenta exceeding $|\delta'|$. 

\noindent
In this situation, the parton that compensates the overall imbalance, 
$\vec{k}_\perp= - \vec{\delta}'$ is radiated off the incoming, quasi-real, parton legs as shown in Fig.~\ref{Figshortkin}. At the same time, the virtual partons after the core splitting ``0''$\to$ ``1''+``2'' enter their respective hard collisions without radiating any offspring on the way.

\bigskip
\begin{figurehere}
\centering
 \includegraphics[width=0.45\textwidth]{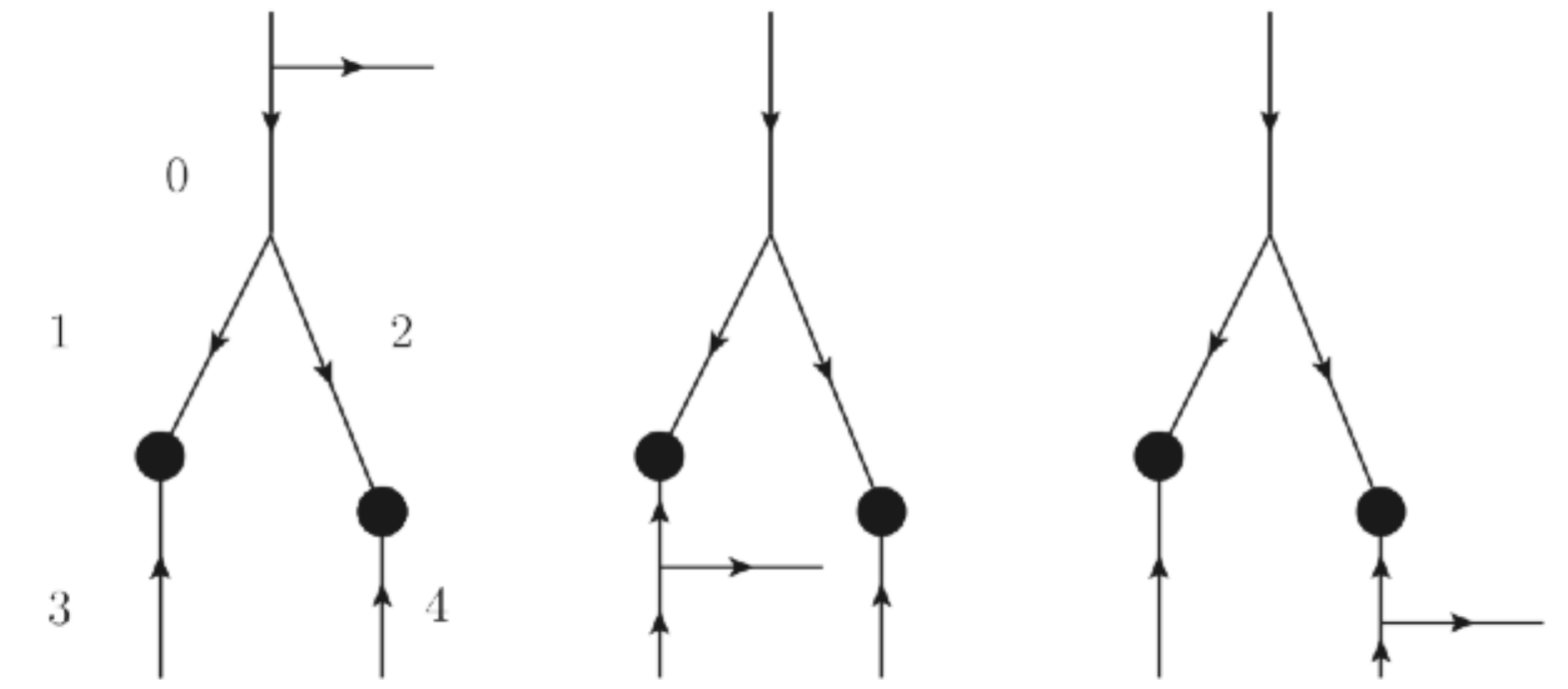}
\caption{\label{Figshortkin} Kinematics of the ``short split'' contribution}
\end{figurehere}

The  $1\to2$ splitting neighbors the hard vertices, therefore the name ``short split'' (aka "endpoint contribution", \cite{BDFS2}).

A closed expression for the differential distribution in jet imbalances due to short split, derived in the leading collinear approximation, is given by Eq.~(27) of \cite{BDFS2}: 
\begin{widetext}
\beq\label{eq:DDT32}
\begin{split}
& \frac{\pi^2\> d\sigma^{\mbox{\scriptsize DPI}}_{\mbox{\scriptsize short}}}{d^2\delta_{13}\, d^2\delta_{24}} \>=\>\>
 \frac {d\sigma_{{\mbox{\scriptsize part}}}
 } {d\hat{t}_1\,d\hat{t}_2}\> \cdot \>   \frac{\as(\delta^2)}{2\pi\, \delta^2} \,  \sum_c  P_{c}^{(1,2)}\!\!
    \left(\frac{x_1}{x_1+x_2}\right) S_1(Q_1^2,\delta^2)\, S_2(Q_2^2,\delta^2) \cr
{ }\quad &\times \frac{d}{d\delta'^2}
 \bigg\{ S_c(\delta^2\!,\delta'^2) \frac{D_{h_1}^{c}(x_1\!+\! x_2,\delta'^2)}{x_1+x_2}
 S_3(Q_1^2 ,\delta'^2) S_4(Q_2^2,\delta'^2) \!
  \int\!\! \frac{d^2\vec{\Delta}}{(2\pi)^2} \, {}_{[2]}\!D_{h_2}(x_3,x_4,\delta'^2\! , \delta'^2;\vec{\Delta} ) \bigg\}
 +  \big\{ h_1 \leftrightarrow h_2 \big\}.
\end{split}
\eeq
\end{widetext}
The short split becomes less important when the scales of the two hard collisions separate.
Indeed, the logarithmic integration over $\delta^2$ is kinematically restricted from above, 
$\delta^2 < \delta^2_{\mbox{\scriptsize{max}}} \simeq \min\{ Q_1^2, Q_2^2\}$. 
As a result, when transverse momenta of jets in one pair much exceed those of the second pair, e.g., 
$ Q_1^2 \gg Q_2^2$ (see \eqref{eq:deltas}),  
the contribution of the short split becomes suppressed as
\[
  \left.  
  {\sigma^{(3\to4)}_{\mbox{\scriptsize short}}}  \right/  {\sigma^{(3\to4)}} \>\propto\> S_1(q_1^2,q_2^2)\, S_3(q_1^2,q_2^2)  \> \ll\> 1
  \quad (Q_1^2\gg Q_2^2).
\]
Here $S_1$ and $S_3$ are the double logarithmic Sudakov form factors of the partons ``1'' and ``3'' that enter the hard interaction with the larger hardness scale. 

Short split induces a strong correlation between jet imbalances which is worth trying to look for experimentally.

The relative weight of the short split depends on the process under consideration. 
For most DPI processes in the kinematical region we have studied, it typically provides 10--15\%\ of the $R$ value. 
However, it happens to be more important when the nature of the process favors parton splitting. 
This is markedly the case of the double Drell-Yan pair production where the short split contribution reaches 30--35\%. 
On the contrary, it turns out to be practically negligible for the same-sign double $W$-meson production (see discussion in the Conclusions section below).

\subsection{Short split in the integrated cross section \label{Sec:ShortInt}}

In the preceding publication we have treated the contribution of the short split to the total DPI cross section as an addition, $1/\sigma_{\mbox{\scriptsize short}}$, 
to the \12\ effective interaction area $1/\sigma_3$, see Eqs.~(28), (32c) of \cite{BDFS2}. 
However, this was not a right thing to do. As it turns out, the short split contribution to the {\em integrated cross section}\/ is already contained in \eqref{eq:eff3}.

To see how this happens, one has to examine the structure of the DDT formula for the \12\ differential cross section \eqref{eq:DDT31} more closely.
The distribution function ${}_{[1]} \!D$ in \eqref{eq:DDT31} in the leading collinear approximation is given by 
Eq.~(18) of \cite{BDFS2}:
\begin{widetext}
\beq\label{eq:termPT}
\begin{split}
{}_{[1]} \!D^{(a,b)}_h(x_1,x_2;   q_1^2,q_2^2; \vec{\Delta}) &= \sum_{a',b',c'} \int^{\min{(q_1^2,q_2^2)}}
\frac{dk^2}{k^2} \frac{\as(k^2)}{2\pi}  \int\!\frac{dy}{y^2}  \>D^{c}_{h}(y,k^2) \cr
& \times  \int\!\frac{dz}{z(1-z)}
 \> P_{c}^{(a',b')}\!\!\left(z\right) \> \cD^{a}_{a'}\left(\frac{x_1}{zy},q_1^2; k^2\right) \cD^{b}_{b'}\left(\frac{x_2}{(1-z)y}, q_2^2 ; k^2\right).
\end{split}
\eeq
\end{widetext}
Here $a,b$ mark the registered partons, and $a',b',c$ are the indices of the partons involved in the splitting 
$c\to a'(z) + b'(1-z)$ with the DGLAP probability $P_{c}^{(a',b')}(z)$. The distribution $D^{c}_h(y, k^2)$ describes the standard probability of finding a parton $c$ inside the incident hadron $h$ at the transverse momentum scale $k^2$, 
and the functions $\cD^i_{i'} (x,q^2; k^2)$ in the second line stand for the distribution of parton $i$, probed at scale $q^2$, in the initial parton $i'$ at lower virtuality scale $k^2$. 

Let us examine the structure of the terms one gets applying two derivatives to \eqref{eq:termPT} substituted into \eqref{eq:DDT31}: 
\beq
  \frac{d}{dq_{1}^2} \frac{d}{dq_{2}^2}   \int^{\min{(q_1^2,q_2^2)}}\!
  \frac{dk^2} {k^2} \as(k^2) {\cal{F}}(k^2;q_1^2,q_2^2) .
\eeq
Taking derivative over $q_i^2$ of the product $D(x,q_i^2)\times S(Q^2,q_i^2)$ corresponds to picking up from the parton chain an accompanying parton $\ell$ (with the largest transverse momentum) which compensates the imbalance, $\ell=-q_i$. 

Apart from logarithmic dependences of $\cD$ and $S$, the derivative in \eqref{eq:DDT31} may act upon the upper limit of the $k^2$integration in \eqref{eq:termPT}.
\begin{subequations}\label{eq:dirs}
\beq\label{eq:dir0}
 \frac{\as}{q_1^2}\frac{\as}{q_2^2}  \int^{\min{(q_1^2,q_2^2)}}\!
  \frac{dk^2} {k^2}\as(k^2) {\cal{F}}''(k^2;q_1^2,q_2^2). 
\eeq  
Differentiating the upper limit of the $k^2$-integral describes another legitimate situation when one of the partons ``1'' and ``2'' does not radiate before entering the hard interaction. In this case the smaller of the jet imbalances is determined by the splitting momentum $k$:
\beq\label{eq:dir1}
 \frac{\as}{q_1^2}\frac{\as}{q_2^2} \left[  \vartheta(q_2^2-q_1^2){\cal{F}}' \big|_{k^2=q_1^2} 
+ \vartheta(q_1^2-q_2^2){\cal{F}}' \big|_{k^2=q_2^2} \right]
\eeq  
Finally, applying {\em both}\/ derivatives to the integration limit \eqref{eq:termPT} gives rise to
\beq\label{eq:dir2}
  \frac{\as}{q_1^2} {\delta(q_1^2-q_2^2)} {\cal{F}}\big|_{k^2=q_1^2=q_2^2}. 
\eeq  
\end{subequations}

\noindent
Contrary to the first two contributions \eqref{eq:dir0} and \eqref{eq:dir1}, the last term \eqref{eq:dir2} 
obviously violates the condition of applicability  \eqref{eq:dlog1} of the DDT formula. 
Instead, in the region $\delta_{13}^2\sim \delta_{24}^2$ it is the short split that contributes in the leading collinear approximation, so that Eq.~\eqref{eq:DDT32}
has to be used to describe the differential spectrum in place of \eqref{eq:dir2}.    

However, as far as the {\em total cross section}\/ is concerned, 
the integrals over imbalances of the short split and of the fake singular term \eqref{eq:dir2} turn out, as by miracle, to be the same. 
Indeed, integrating the short split \eqref{eq:DDT32} over $\delta'^2$ up to $\delta^2$, we get
\begin{widetext}
\beq \label{eq:Xshort}
  \sum_c  P_c^{(1,2)}\left(\frac{x_1}{x_1+x_2}\right)
 \int^{ \min\{Q_1^2,Q_2^2\}} 
 \frac{d\delta^2}{\delta^2}\frac{\alpha_s(\delta^2)}{2\pi} 
 \frac{D^c_{h_1}(x_1+x_2,\delta^2)}{x_1+x_2}  \prod_{i=1}^4 S_i \cdot 
  \int\frac{d^2 {\Delta}}{(2\pi)^2} { } _{[2]} D
 _{h_2}(x_3,x_4, \delta^2,\delta^2;\vec\Delta),
 \eeq
\end{widetext}
where 
$ \prod_{i=1}^4 S_i \equiv  S_1(Q_1^2,\delta^2)S_2(Q_2^2,\delta^2)S_3(Q_1^2,\delta^2)S_4(Q_2^2,\delta^2)$. 

On the other side, taking the integrand of ${}_{[1]} D$ \eqref{eq:termPT} in the point $q_1^2=q_2^2=k^2$, and using 
\begin{eqnarray*}
\left.  \cD^{a}_{a'}\bigg(\frac{x_1}{zy},q_1^2; k^2\bigg)\right|_{q_1^2=k^2} &=& \delta_{a'}^{a}\delta\bigg(1-\frac{x_1}{zy}\bigg), \\
\left. \cD^{b}_{b'}\bigg(\frac{x_2}{(1-z)y}, q_2^2 ; k^2\bigg)\right|_{q_2^2=k^2} &=& \delta_{b'}^{b}\delta\bigg(1-\frac{x_2}{(1-z)y}\bigg),
\end{eqnarray*}
we evaluate the momentum integrals,
\[
 \int\! \frac{dy}{y^2}   \int\!\! \frac{dz}{z(1\!-\!z)} \delta\bigg(1-\frac{x_1}{zy}\bigg) \delta\bigg(1-\frac{x_2}{(1\!-\!z)y}\bigg) 
 = \frac1{x_1+x_2},
\]
to arrive at the very same expression \eqref{eq:Xshort}.

It is worth noticing that this correspondence does not depend on the precise form of the upper integration limit  in \eqref{eq:termPT}. The result does not change, within the leading logarithmic accuracy, if one replaces a sharp $\vartheta$-function cut by a smooth damping factor that cuts the logarithmic $k^2$ integration at $k^2\sim\min\{q_1^2,q_2^2\}$.

\medskip

Thus, for the integrated DPI cross section we obtain two contributions to the effective interaction area:
\begin{widetext}
\begin{subequations}\label{eq:eff43}
\beeq
\label{eq:eff4}
\frac{\prod_{i=1}^4 \! D(x_i)}{\sigma_{4}} &\!\!\!=\!\!\!\!&  
\int\! \frac{d^2\vec{\Delta}}{(2\pi)^2} \> _{[2]}D_{h_1}
(x_1,x_2, Q_1^2,Q_2^2;\vec\Delta) 
\> {}_{[2]}D_{h_2}
(x_3,x_4, Q_1^2,Q_2^2; -\vec\Delta), \\
\label{eq:eff3}
\frac{\prod_{i=1}^4 \! D(x_i)}{\sigma_{3}} &\!\!\!=\!\!\!\!\!&  
\int\!\! \frac{d^2\vec{\Delta}}{(2\pi)^2}  \! \bigg[ {}_{[2]}\!D_{h_1}\!
(x_1,x_2, Q_1^2,Q_2^2; \!\vec\Delta) {}_{[1]}\!D_{h_2}\!
(x_3,x_4, Q_1^2,Q_2^2) 
+\! {} _{[1]}\!D_{h_1}\!
(x_1,x_2, Q_1^2,Q_2^2) {}_{[2]}\!D_{h_2}\!
(x_3,x_4, Q_1^2,Q_2^2;  \!\vec\Delta)\!\bigg]\!. \qquad {  }
\eeeq
\end{subequations} 
\end{widetext}

Let us stress in conclusion that a compact and intuitively clear expression containing the product of the $_2$GPDs ${}_{[2]}\mbox{D}$ and ${}_{[1]}\mbox{D}$ in \eqref{eq:eff3} applies only to the {\em integrated}\/ \12\ cross section. 

When addressing the {\em differential}\/ distributions, one has to employ the ``DDT-like formula'' \eqref{eq:DDT31} in the region of strongly ordered transverse momenta \eqref{eq:dlog1}, 
and a quite different expression \eqref{eq:DDT32} in the kinematical region of nearly opposite jet pair imbalances \eqref{eq:dlog2}.  

\section{Modeling ${}_{[2]}\mbox{D}$ \label{SEC:Model}}
To proceed with quantitative estimates, one needs a model for the non-perturbative two-parton distributions in a proton. 

A priori, we know next to nothing about them.
The first natural step to take is an {\em approximation of independent partons}.  
It allows one to relate $_2$GPD with known objects, namely~\cite{BDFS1}
\begin{subequations}\label{eq:model}
\beq\label{eq:DGG}
{}_{[2]}D(x_1,x_2, Q_1^2,Q_2^2;\Delta) \simeq G(x_1,Q_1^2;\Delta^2) G(x_2, Q_2^2;\Delta^2).
\eeq
Here $G$ is the non-forward parton correlator (known as generalized parton distribution, GPD) 
that determines, e.g., hard vector meson production at HERA
and which enter in our case in the diagonal kinematics in $x$ ($x_1=x_1'$,  see Fig.~\ref{FigGPD}). 

\begin{figurehere}
\centering
 \includegraphics[width=0.35\textwidth]{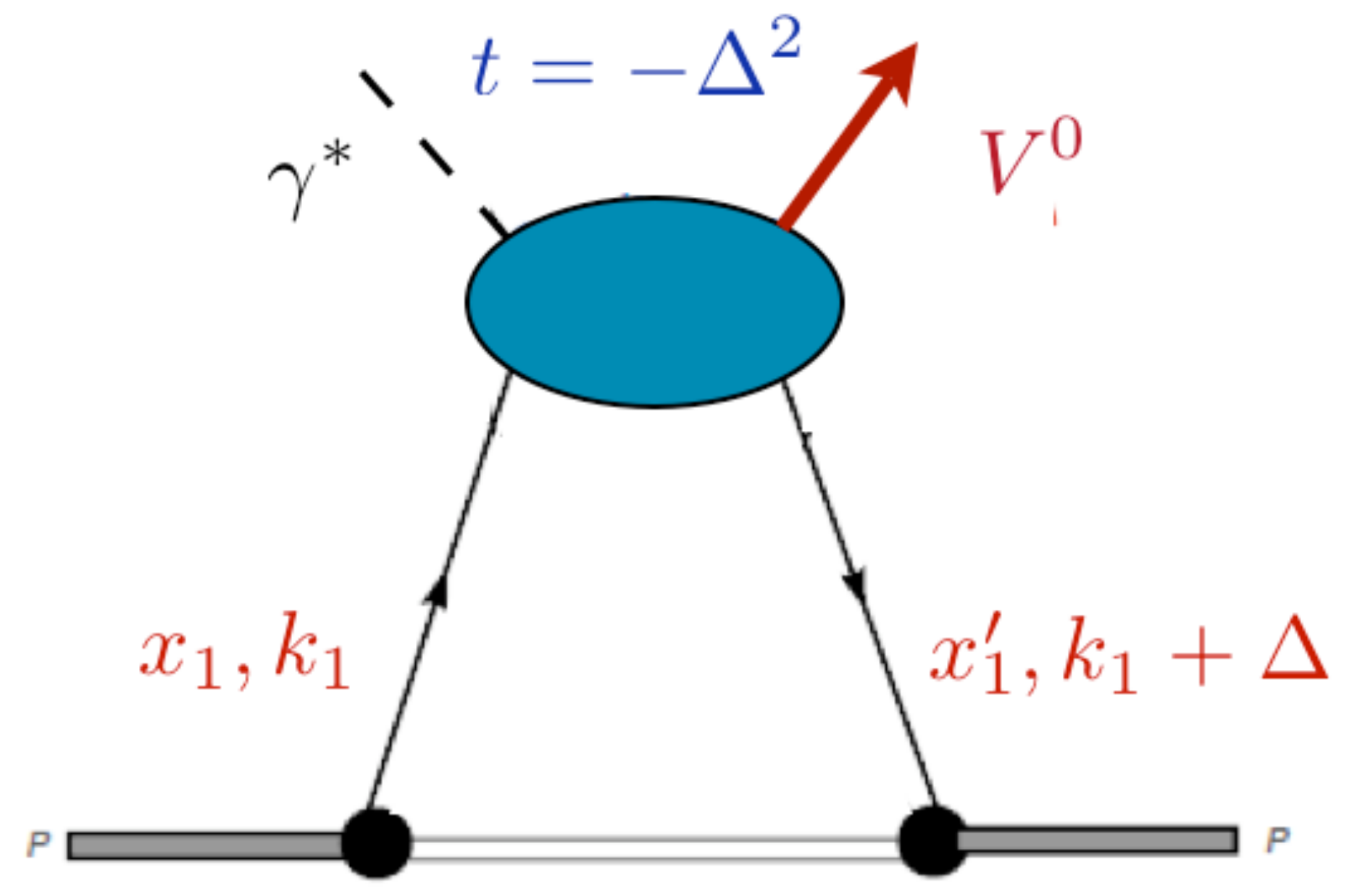}
\caption{\label{FigGPD} GPD in the vector meson electroproduction amplitude}
\end{figurehere}

\noindent
The modeling by \eqref{eq:DGG} is not perfect. 
First of all, it does not respect an obvious restriction $D(x_1+x_2> 1)=0$. So, $x_i$ have to be taken not too large (say, $x_i<0.5$).  
Actually, $x_i$ must be taken even smaller. The GPD  in Fig.~\ref{FigGPD} is an elastic amplitude, while the corresponding block in the DPI represents the inclusive cross section (the cut-through amplitude).  
For the analogy to hold, the interaction amplitude has to be close to imaginary. This condition calls for $x_i< 0.1$. 

On the other hand, $x_i$ should not be {\em too small}\/ to stay away from the region of the Regge-Gribov phenomena where there are serious reasons for parton correlations to be present at the non-perturbative level (see discussion in \cite{BDFS3}).  

Thus, we fix the domain of applicability of the model \eqref{eq:DGG} for $_2\mbox{GPD}$ 
as $10^{-1}> x_i > 10^{-3}$. 

The GPD, in its turn, can be modeled as
\beq\label{eq:GDF}
 G(x_1,Q_1^2;\Delta^2) \>\simeq\> D(x_1, Q_1^2) \times F_{2g}(\Delta^2) ,
\eeq
with $D$ the usual one-parton distribution determining DIS structure functions and $F$ the so-called two-gluon form factor of the hadron. 
The latter is a non-perturbative object; it falls fast with the ``momentum transfer'' $\Delta^2$. 
This form factor can be parametrized differently. For example, by a dipole formula
\beq\label{eq:Fdipole}
     F_{2g}(\Delta^2) = \left( 1+ \frac{\Delta^2}{m_g^2}\right)^{-2},
\eeq 
\end{subequations}
where an effective parameter $m_g^2$ extracted from the FNAL and HERA $J/\psi$ exclusive photoproduction data 
lies in the ballpark of $m_g^2(x\sim 0.03, Q^2\sim 3\,\GeV^2)\simeq 1.1\,  \GeV^2$  
and decreases with further decrease of $x$ \cite{FS}.

Substituting \eqref{eq:model} into \eqref{eq:XsS} gives 
\beq\label{eq:seff1}
{\sigma_{\eff}}^{-1} \>=\>
\int\! \frac{d^2\vec{\Delta}}{(2\pi)^2} \> F_{gg}^4(\Delta^2) \>=\> \frac{m_g^2}{28\pi} , \quad  
\sigma_{\eff} \simeq 32\,\rm{mb}. 
\eeq
It is about a factor of two larger than the value measured by the Tevatron experiments \cite{Tevatron1,Tevatron2}. 

pQCD induced parton correlations (\12\ DPI processes) are capable of explaining this discrepancy. 
	
\medskip

Turning to the \12\ term, we neglect a mild logarithmic $\Delta$-dependence of $_{[1]}D$ in \eqref{eq:eff3} and use the model \eqref{eq:model} for $_{[2]}D$  to obtain 
\beq
 {\sigma_{3}}^{-1}  \>\simeq\> 
 \frac73\cdot \left[ \frac{_{[1]}D(x_1,x_2)}{D(x_1)D(x_2)} + \frac{_{[1]}D(x_3,x_4)}{D(x_3)D(x_4)} \right] \times {\sigma_{4}}^{-1} , 
\eeq
where we substituted the value of the integral (cf.\ \eqref{eq:seff1})
\[
     \int\! \frac{d^2\vec{\Delta}}{(2\pi)^2} \> F_{gg}^2(\Delta^2) \>=\>\frac{m_g^2}{12\pi}. 
\]
We will parametrize the result in terms of the ratio 
 \beq\label{eq:Rdef}
      R\>\equiv\>  \frac{\sigma_{\mbox{\scriptsize \12}}}{\sigma_{\mbox{\scriptsize \22}}} \>=\> \frac{\sigma_4}{\sigma_3}. 
 \eeq
For the effective interaction area, 
\beq\label{eq:Sfull}
 \sigma_\eff^{-1} \>=\> \sigma_{4}^{-1} + \sigma_{3}^{-1},  
\eeq
we then have
\beq\label{eq:Smodel}
   \sigma_\eff \>=\> \frac{28\pi}{m_g^2}\cdot\frac1{1+R} \>\>\simeq\>\> \frac{35\,\rm{mb}}{m_g^2\,[\GeV] }\cdot \frac1{1+R} 
   \>\>\simeq\>\> \frac{32\,\rm{mb}}{1+R}
\eeq
(the phenomenological value $m_g^2=1.1\,\GeV^2$ was used). 

Within the framework of the NP two-parton correlations model \eqref{eq:DGG}, 
there is but one free parameter $Q_0^2$. 
The DPI theory applies to various processes and holds in a range of energies and different kinematical regions. 
Therefore, having fixed the $Q_0^2$ value, say, from the Tevatron data, one can consider all other applications (in particular, to LHC processes) as parameter-free theoretical predictions.  

\section{Numerical results \label{SEC:Numerics}}

\subsection{Calculation framework}

In numerical calculations we used the GRV92 parametrisation of gluon and quark parton distributions in the proton \cite{GRV}.  
We have checked that using more advanced GRV98 and CTEQ6L parametrisations does not change the numerical results.  
The explicit GRV92 parametrisation is speed efficient and allows one to start the PT  
evolution from rather small virtuality scales. 
The combination $(Q_0^2+\Delta^2)$ was used as the lower cutoff for logarithmic transverse momentum integrals in the parton evolution, 
which induced a mild (logarithmic) $\Delta$-dependence on top of the relevant power of the two-gluon form factor $F_{2g}(\Delta^2)$.
   
To quantify the role of the \12\  DPI subprocesses, we calculated the ratio $R$ \eqref{eq:Rdef}
in the kinematical region $10^{-3} < x_i < 10^{-1}$ for Tevatron ($\sqrt{s}=1.8 \div 1.96 \, \mbox{TeV}$) and LHC energies ($\sqrt{s}= 7\, \mbox{TeV}$). 
We chose to consider three types of  ensembles of colliding partons: 
\begin{enumerate}
\item
 $u(\bar{u})$ quark and three gluons which is relevant for ``photon plus 3 jets'' CDF and D0 experiments,
\item
four gluons (two pairs of hadron jets),
\item
$u\bar{d}$ plus two gluons, illustrating  $W^+jj$ production.
\item
$u\bar{d}$ plus $d\bar{u}$, corresponding to the $W^+W^-$ channel.
\end{enumerate}   
\medskip

 \subsection{Perturbative \12 correlation at the Tevatron.}

\subsubsection{CDF experiment}

In Fig.~\ref{FigCDF10} we show the profile of the  \12\ to \22\ ratio $R$ for the $\gamma+3$jets process
in the kinematical domain of the CDF experiment \cite{Tevatron1}.   
The calculation was performed for the dominant ``Compton scattering'' channel of the photon production: $g(x_2)+u(\bar{u})(x_4)\to \gamma + u(\bar{u})$.  
The longitudinal momentum fractions of two gluons producing second pair of jets are $x_1$ and $x_3$. 
The typical transverse momenta were taken to be $p_{\perp 1,3}\simeq 5\,\GeV$ for the jet pair, 
and $p_{\perp 2,4}\simeq 20\, \GeV$ for the photon--jet system. 
In Fig.~\ref{FigCDF10} $R$ is displayed as a function of rapidities of the photon--jet,  $\eta_2 = \half\ln(x_2/x_4)$, and the 2-jet system, 
$\eta_1 = \half\ln(x_1/x_3)$. 

\begin{widetext}
\begin{figurehere}
\centering
\begin{minipage}{0.45\textwidth}
 \includegraphics[width=\textwidth]{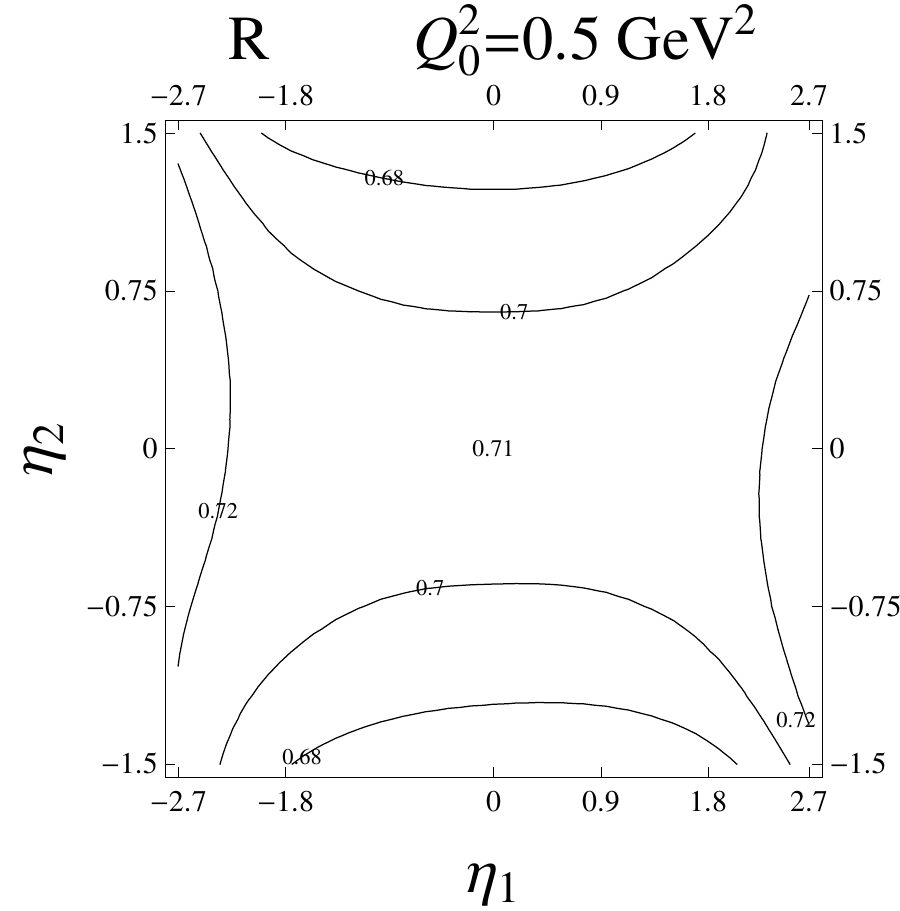}
\end{minipage}
\hfill 
\begin{minipage}{0.45\textwidth}
\centering
 \includegraphics[width=\textwidth]{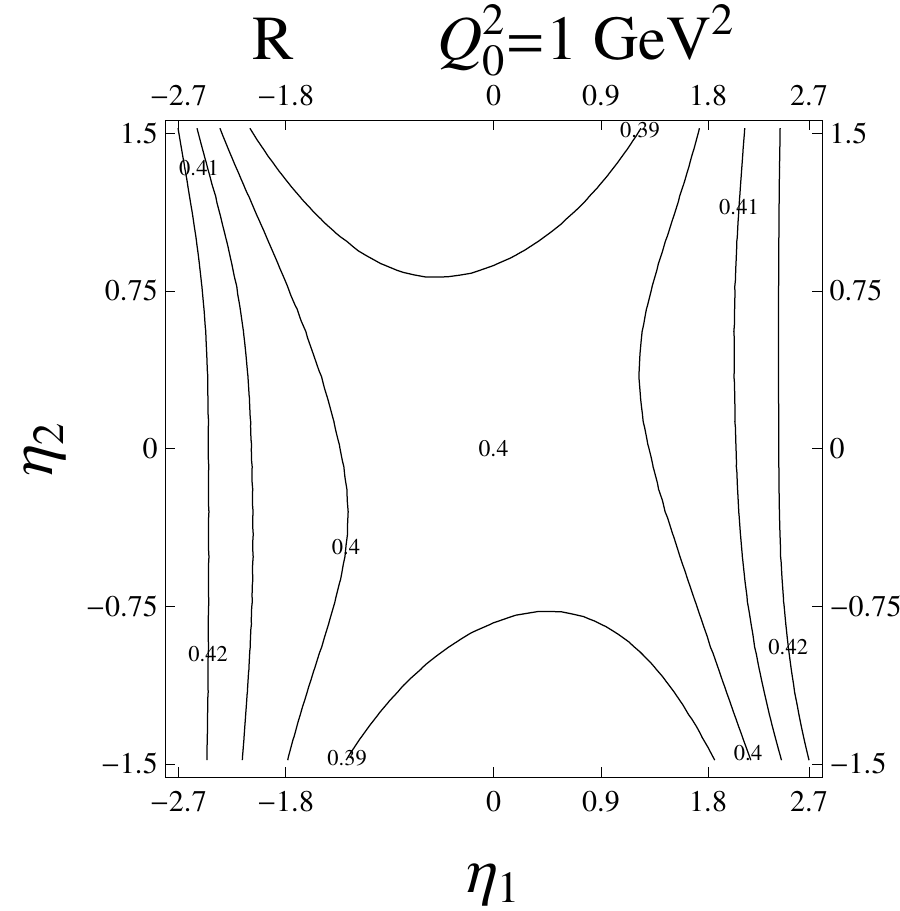}
\end{minipage}
\caption{\label{FigCDF10} The  \12/\22\ ratio \eqref{eq:Rdef} in the CDF kinematics
for the process $p\bar p \to \gamma + 3\, \mbox{jets}  + X$.} 
\end{figurehere}
\end{widetext}
We observe that the enhancement factor lies in the ballpark of $1+R\simeq 1.5\div 1.8$. Processed through Eq.~\eqref{eq:Smodel}, 
it translates into $\sigma_\eff \simeq 18\div21$~mb. This expectation has to be compared with the CDF finding 
$\sigma_\eff =14.5 \pm 1.7\>^{+1.7}_{-2.3}\>\mbox{mb}$. 
A recent reanalysis of the CDF data  points at an even small value: $\sigma_\eff =12.0\pm 1.4\>^{+1.3}_{-1.5}\>\mbox{mb}$,  \cite{Sjodmok}.

In our previous report \cite{BDFS3} we have included a plot of the $x$-dependence of $R$ for central production at Tevatron and LHC energies. 
That plot turned out to be confusing: a rather sharp $x$-dependence it has demonstrated seemed to contradict the CDF findings of approximate constancy of $\sigma_\eff$. 
In fact, this variance with $x$ was in the major part resulting from the kinematical link between $x$ and $Q^2$ for a given collision energy ($Q^2=x^2s/4$).  

The results of numerical calculation for a fixed hardness $Q^2$ shown in Fig.~\ref{FigCDF10} 
for the CDF kinematics exhibit a very mild $x$-dependence of the $R$ factor and thus of the $\sigma_\eff$. 

\subsubsection{D0 experiment}

\begin{figure}[h]  
	\centering
	 \includegraphics[width=0.4\textwidth]{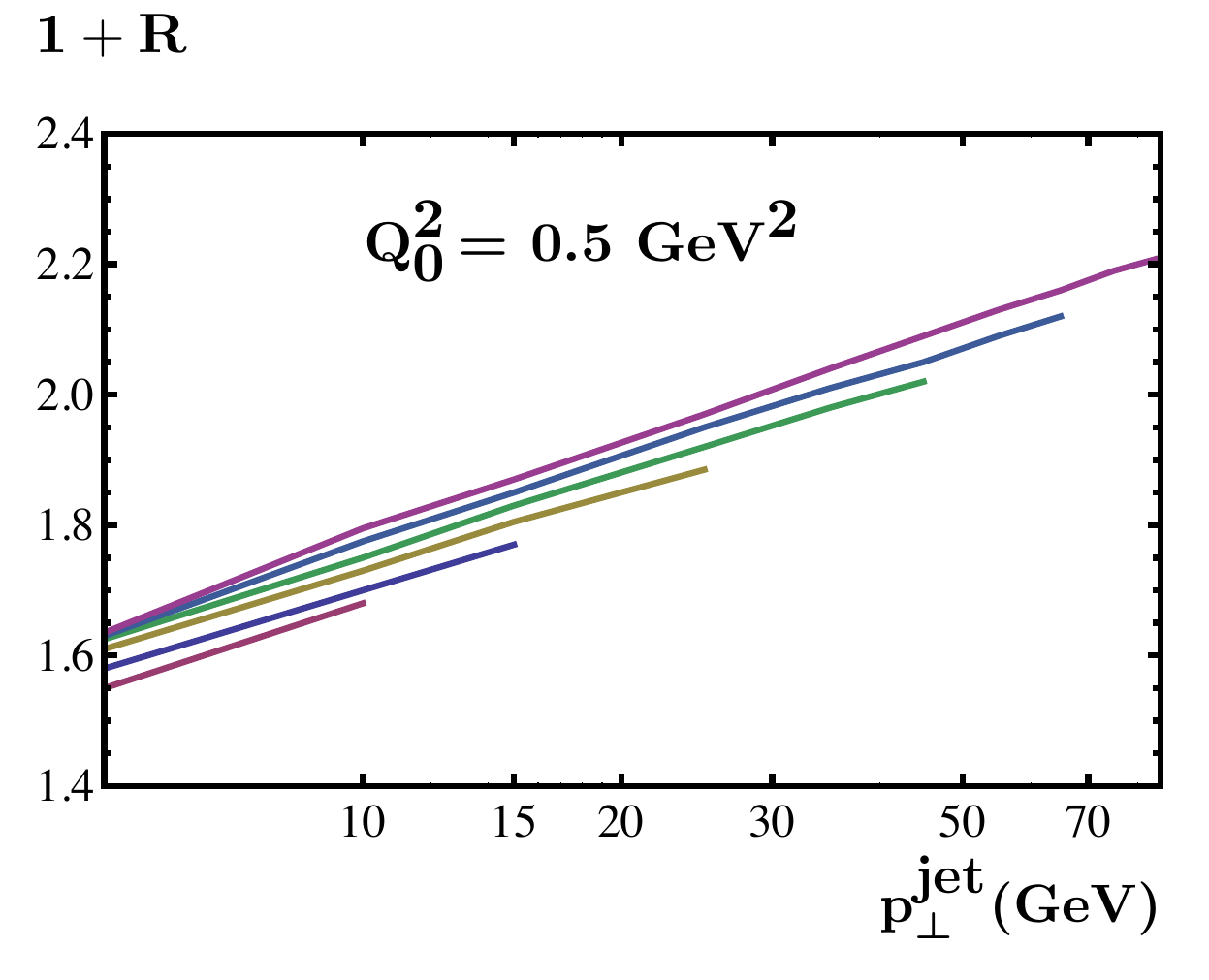}  
	 \vspace{-5mm}
\caption{\label{FigD0pt1} Central rapidity photon+3 jets production in $u$($\bar{u}$)--gluon collisions
in the D0 kinematics. }
\end{figure}

\begin{figure}[h]  
	\centering
	 \includegraphics[width=0.4\textwidth]{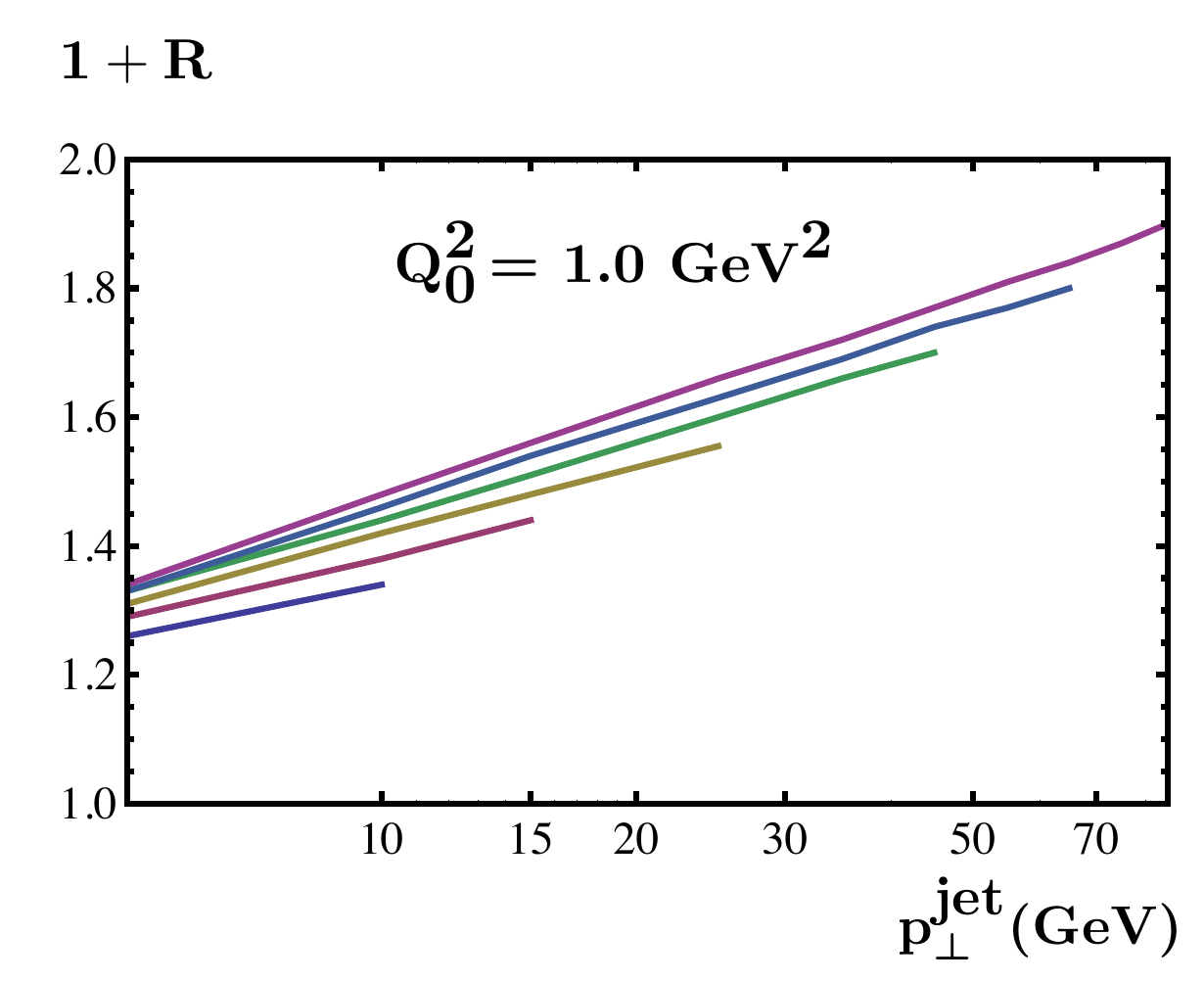}  
	 \vspace{-5mm}
\caption{\label{FigD0pt2} Same as Fig.~\ref{FigD0pt1} for $Q_0^2=1\GeV^2$. }
\end{figure}

The ratio $R$ is practically constant in the kinematical domain of the D0 experiment on photon+3 jets production \cite{Tevatron2,Tevatron3} and is very similar to that of the CDF experiment shown above in Fig.~\ref{FigCDF10}. 
So, for the D0 kinematics we instead display in Figs.~\ref{FigD0pt1},  \ref{FigD0pt2} 
the enhancement factor $1\!+\!R$ in dependence on $p_\perp$ of the secondary jet pair
for  photon transverse momenta 10, 20, 30, 50, 70, and 90 \GeV (from bottom to top).  

 \clearpage

\begin{figure}
\centering
\begin{minipage}{0.45\textwidth}
 \includegraphics[width=0.95\textwidth]{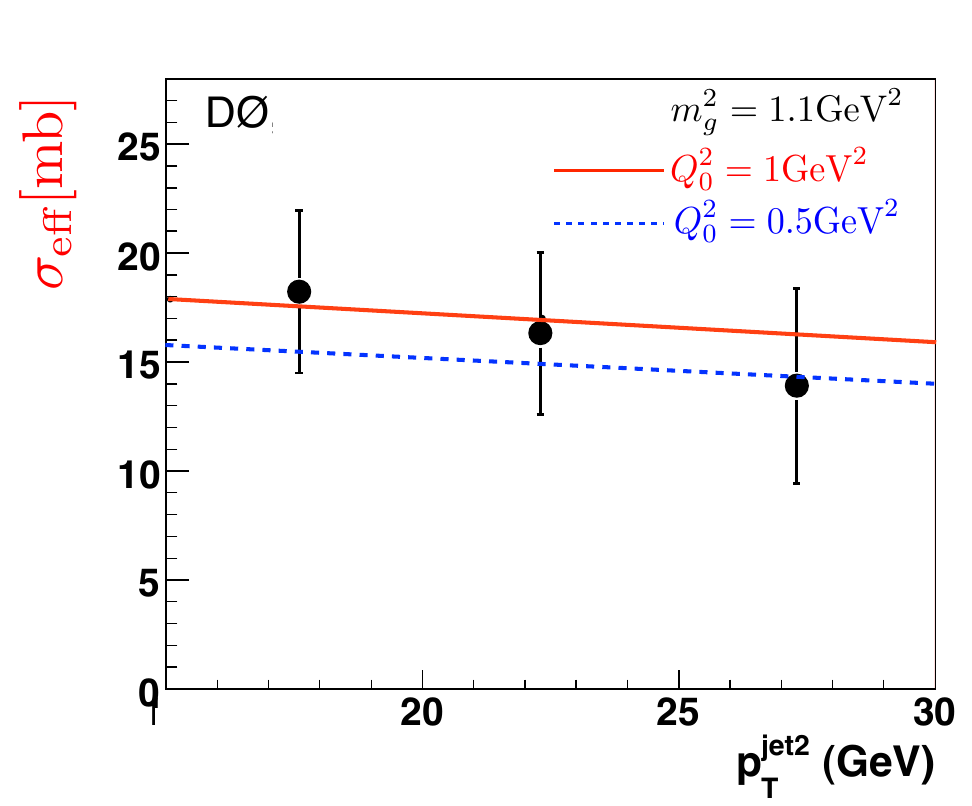}
 	 \vspace{-5mm}
\caption{\label{FigD0Seff} $\sigma_\eff$ as a function of the hardness of the second jet in the kinematics of the D0 experiment for $p_{\perp\gamma}=70\,\GeV$.}
\end{minipage}
\end{figure}

The corresponding prediction for $\sigma_\eff$ is shown in Fig.~\ref{FigD0Seff} in comparison with the D0 findings. 
Both the absolute value and (a hint at) the $p_\perp$-trend look satisfactory.

\subsection{LHC energies}

In Fig.~\ref{FigLHCgg05} we show the \12\ to \22\ ratio for production of two pairs of back-to-back jets with transverse momenta $50\,\GeV$ produced in collision of gluons at the LHC energy $\sqrt{s}=$ 7\, TeV.

\begin{widetext}
\begin{figurehere}
\begin{minipage}{0.43\textwidth}
\includegraphics[width=\textwidth]{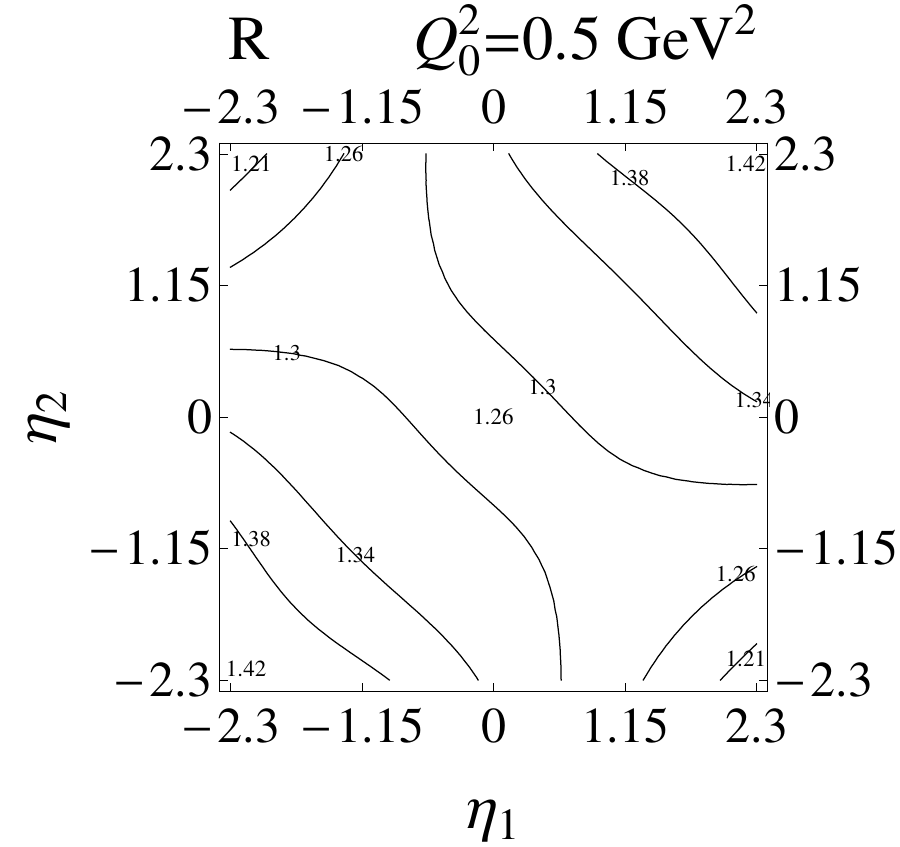}
\end{minipage}
\hfill 
\begin{minipage}{0.43\textwidth}
\centering
\includegraphics[width=\textwidth]{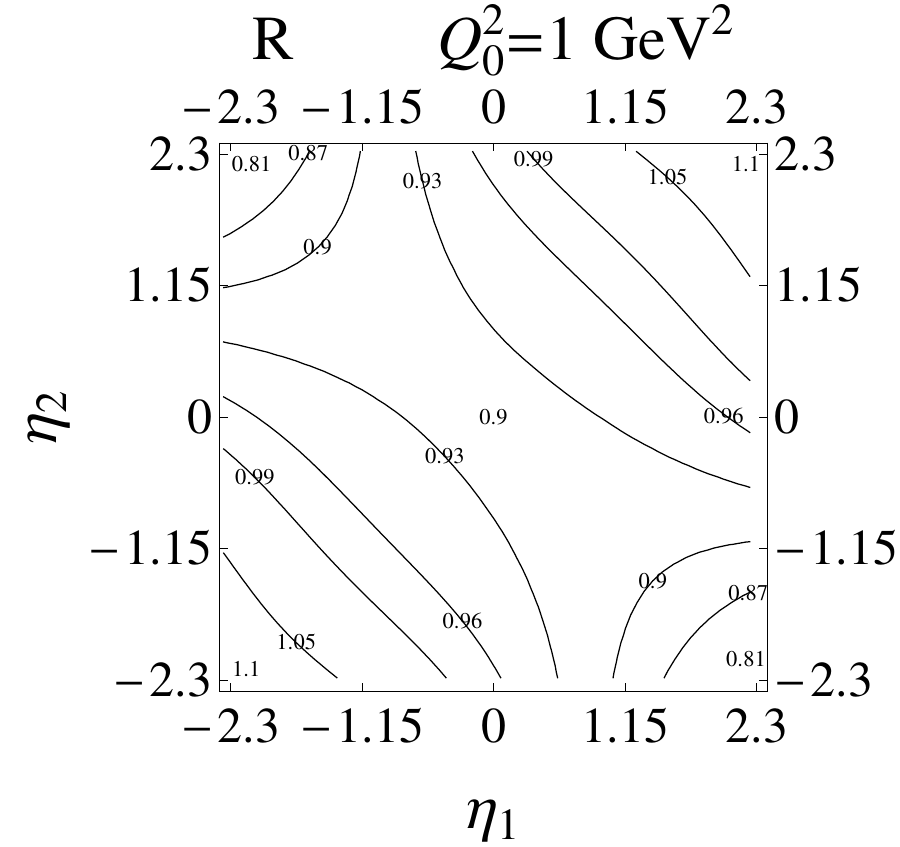}
\end{minipage}
	 \vspace{-2mm}
\caption{\label{FigLHCgg05} Rapidity dependence of the $R$ factor for two pairs of $p_\perp=50\,\GeV$ jets produced in gluon-gluon collisions}
\end{figurehere}
\end{widetext}

\begin{figure}[b]
\centering
\begin{minipage}{0.43\textwidth}
  \includegraphics[width=0.9\textwidth]{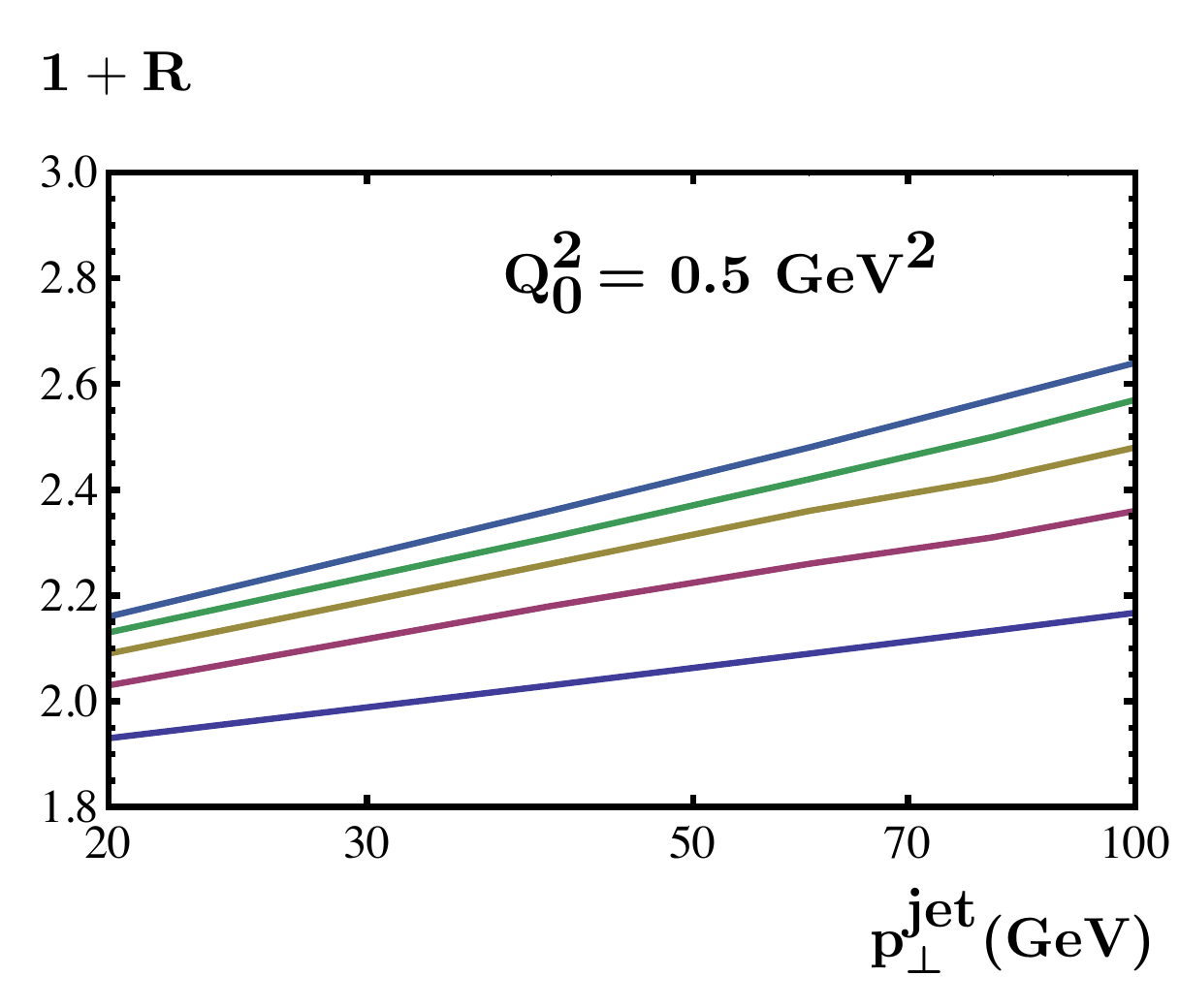}  
  	 \vspace{-7mm}
\caption{\label{FigJET} $1+R$ for two dijets at LHC: $p_{\perp1}=$ 20, 40, 60, 80, 100 \GeV (from bottom to top).}
\end{minipage}
\end{figure}

Dependence on the hardness parameters of the DPI process of double gluon--gluon collisions is illustrated in Fig.~\ref{FigJET}. 
For the sake of illustration, we have chosen  the value of the PT cutoff parameter $Q_0^2=0.5\,\GeV^2$, and calculated the enhancement factor $1+R$ for five values of the transverse momenta of the jets in one pair, $p_{\perp1}=$ 20, 40, 60, 80, 100 \GeV. 

Fig.~\ref{FigJET} demonstrates its dependence on the transverse momenta of jets in the second pair, $ p_\perp \equiv p_{\perp2}$. 

We observe that within the chosen range, $R$ increases by about 15--25\%\ with increase of the hardness of one of the jet pairs. This corresponds to approximately 10\%\ drop in $\sigma_\eff$. 

\clearpage

Finally, in Fig.~\ref{FigLHCwjj05} we show the rapidity profile of the $R$ ratio for the process of production of the vector boson, 
$u\bar{d}\to W^+$, accompanied by an additional pair of (nearly back-to-back) jets with transverse momenta $p_\perp = 30\,\GeV$ produces in a gluon--gluon collision.  
 \vfill
 
\begin{widetext}
	\begin{figurehere}
	\centering
	\begin{minipage}{0.43\textwidth}
\includegraphics[width=\textwidth]{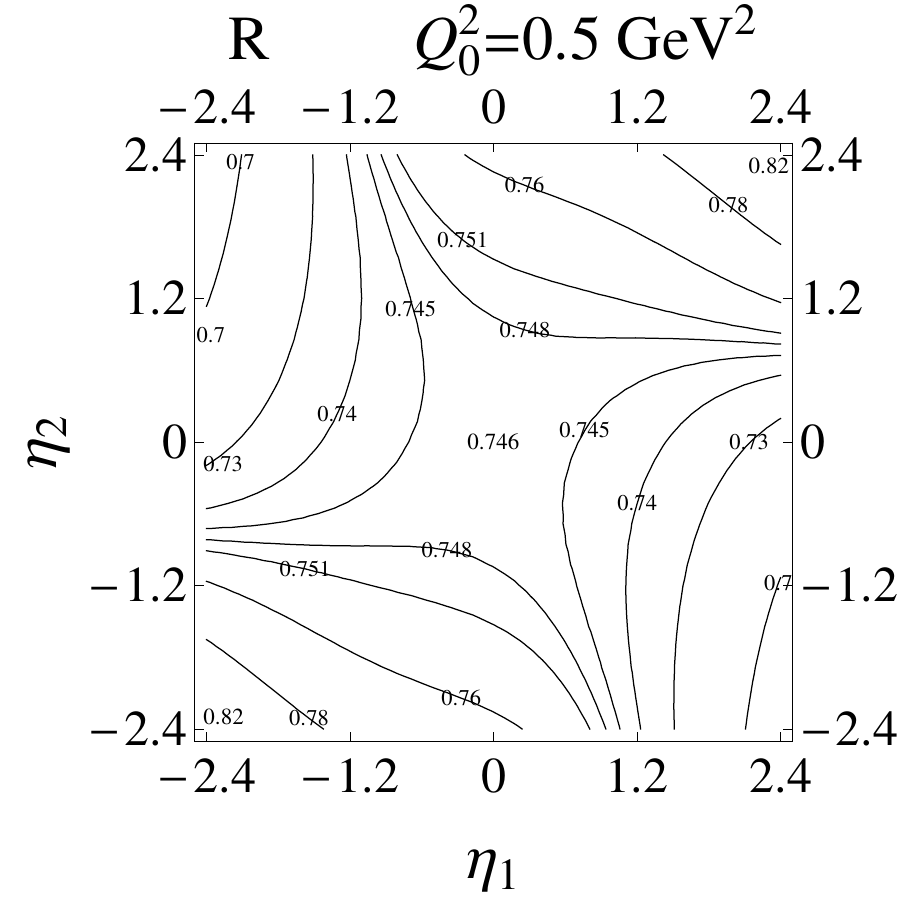}
	\end{minipage}
	\hfill 
	\begin{minipage}{0.43\textwidth}
	\centering
\includegraphics[width=\textwidth]{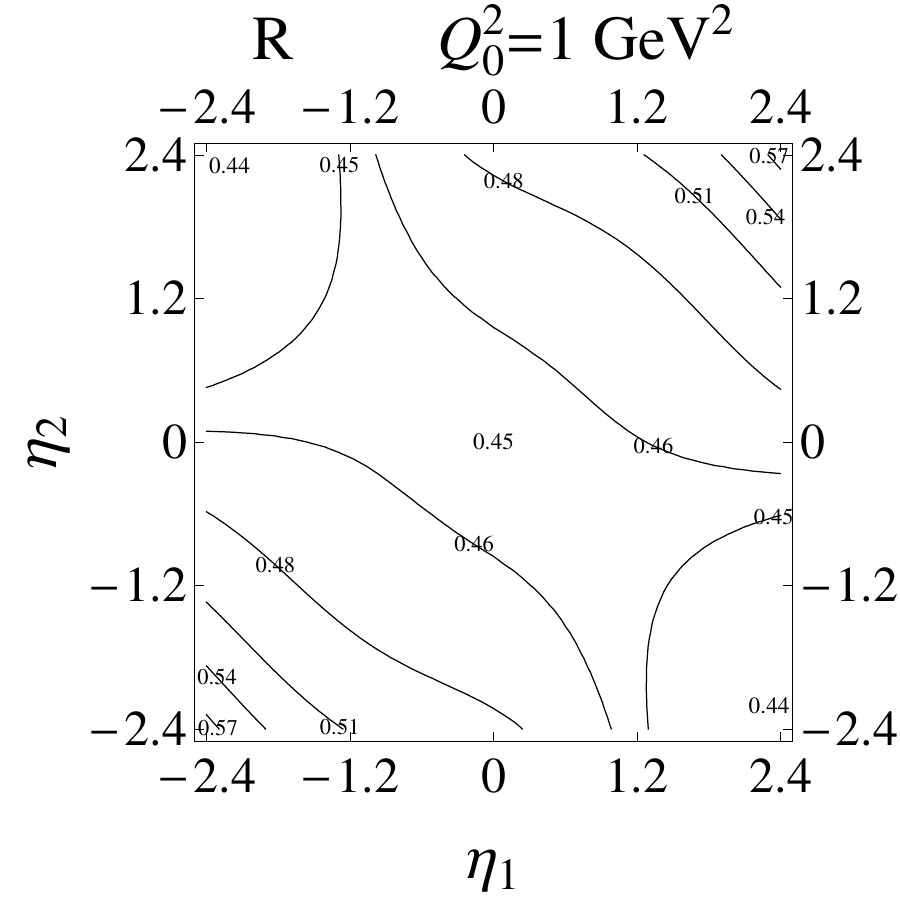}
	\end{minipage}
	 \vspace{-2mm}
	\caption{\label{FigLHCwjj05} Ratio $R$ for production of $W$ plus a pair of $p_\perp \simeq 30\,\GeV$ gluon jets}
	\end{figurehere}
\end{widetext}
	
It is interesting to notice that the effect of perturbatively induced parton--parton correlations 
is maximal for equal rapidities of the $W$ and the jet pair, and diminishes when they separate.  
This feature is more pronounced when the cutoff parameter $Q_0^2$ is taken larger, so that the PT correlation becomes smaller and, at the same time, exhibits a stronger rapidity dependence. 

The recent ATLAS study \cite{Aad:2013bjm} reported for this process the value
$\sigma_{\eff} = 15 \pm 3\>^{+5}_{-3}$~mb
which is consistent with the expected enhancement due to contribution of the \12\ DPI channel, see \eqref{eq:Rdef}. 

Fig.~\ref{FigseffWjj} shows a slight variation of $\sigma_\eff$ with the jet scale in $Wjj$ production. 

\begin{figurehere}
\centering
 \includegraphics[width=0.4\textwidth]{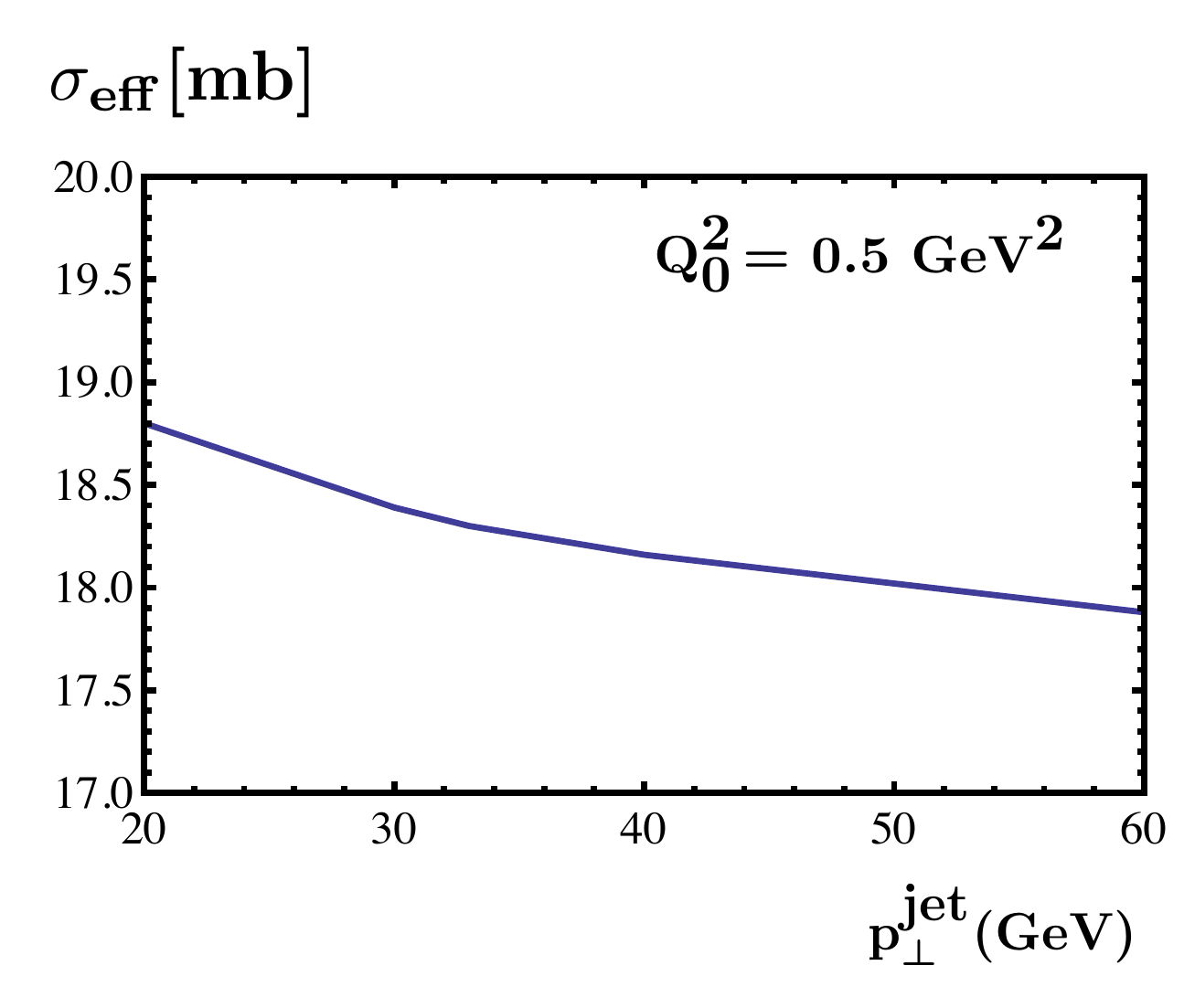}  
\vspace{-5mm}
\caption{\label{FigseffWjj} 	 
$\sigma_\eff$ for the $Wjj$ process at LHC energy as a function of jet transverse momentum.}
\end{figurehere}

\subsection{$Q_0$-dependence}
The dependence of the enhancement factor $1+R$ on the $Q_0$ parameter is shown for the typical kinematics of the CDF photon+3 jets experiment in Fig.~\ref{Figjets50Q0MD1}, and for central production of two pairs of $p_\perp \simeq 50\, \GeV$ jets at the LHC ($\eta_1=\eta_2$) in Fig.~\ref{Figjets50Q0MD2}.

\medskip

\begin{figurehere}
\centering
 \includegraphics[width=0.4\textwidth]{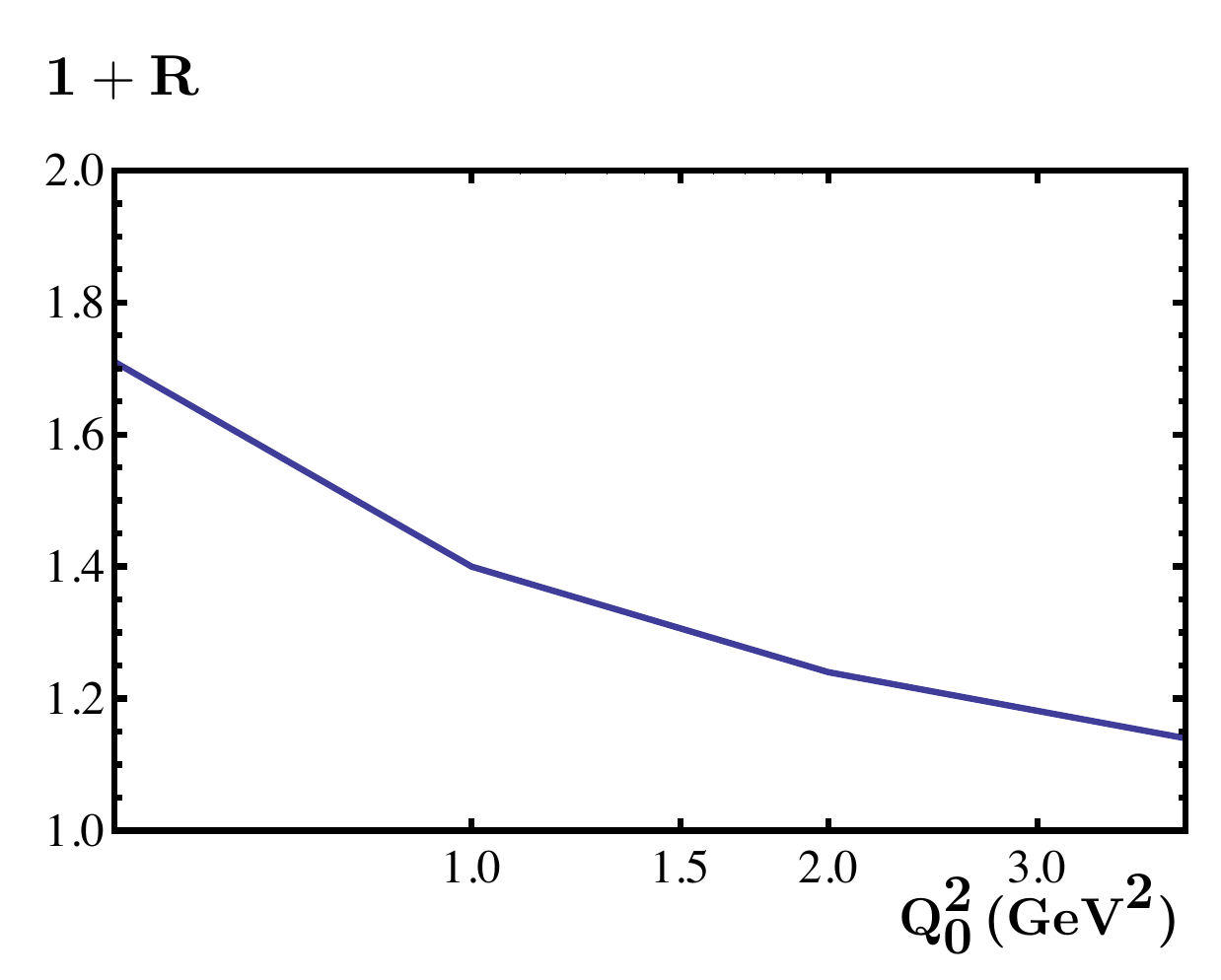} 
 \vspace{-5mm}
\caption{\label{Figjets50Q0MD1} 
$Q_0$-dependence of the enhancement factor $1+R$ for the CDF $\gamma+3\mbox{jets}$ experiment.}
\end{figurehere}

\begin{figurehere}
\centering
	 \includegraphics[width=0.4\textwidth]{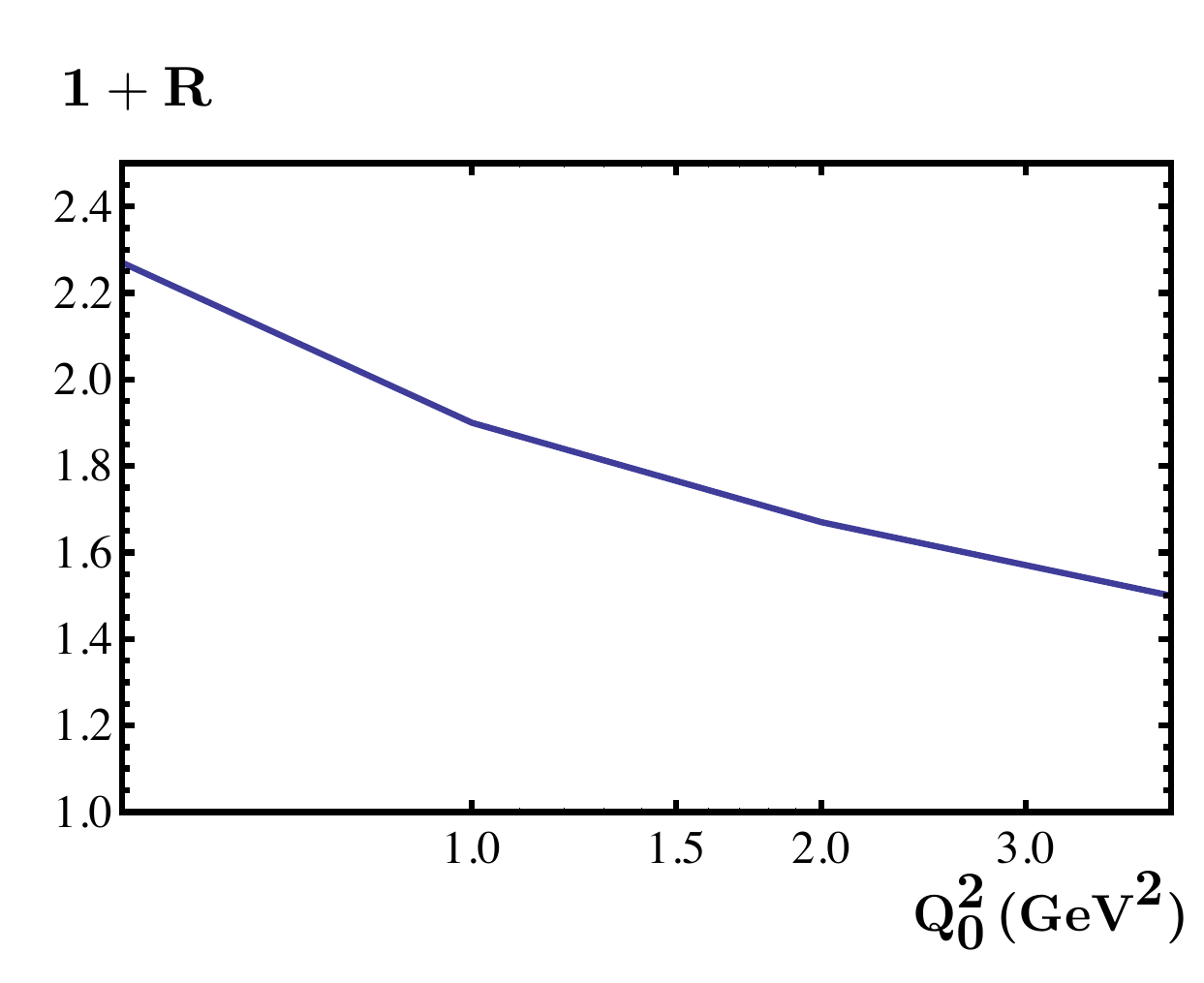}  
\caption{\label{Figjets50Q0MD2} 
$Q_0$-dependence of the enhancement factor $1+R$ for $p_\perp\simeq 50\, \GeV$ LHC dijets.}
\end{figurehere}

\section{Conclusions and Discussion\label{SEC:CONC}}

In the previous paper \cite{BDFS2} we have analyzed the perturbative correlation that arises due to \12\ splitting of the parton in one of the colliding hadrons and derived the corresponding expressions in the leading collinear approximation of the pQCD.  
Here we presented results of numerical evaluation of this contribution to the DPI cross section 
measured at the Tevatron and found the theoretical results to be consistent with the data for the value of the model parameter $Q_0^2\simeq 0.5\, \GeV^2$.   
With $Q_0^2$ fixed, theoretical expectations for certain exemplary DPI processes 
at LHC energies become parameter-free predictions.  

Theoretical derivation of the effective interaction area $\sigma_\eff$ (``effective cross section'') relied on certain assumptions and approximations. 
Our approach to perturbative QCD effects in DPI developed in \cite{BDFS2} was essentially probabilistic. 
In particular, we did not discuss the issue of possible interference between \12\ and \22 two-parton amplitudes. One can argue that such eventuality should be strongly suppressed. 
Indeed, spatial properties of accompanying radiation produced by so different configurations make them unlikely to interfere, since in the \22\ mechanism a typical transverse distance between two partons from the hadron w.f.\ is of order of the hadron size, 
while in the \12\ case it is much smaller and is determined by a hard scale. 
Moreover, we disregarded potential contributions from non-diagonal interference diagrams 
that are due to crosstalk between partons in the amplitude and the amplitude conjugated. Such contributions are absent in \22\ collisions but emerge
specifically in the \12\ DPI process in which the two partons from the hadron wave function are relatively close to one another in the impact parameter plane \cite{Gaunt13}.  

Our prediction for $\sigma_\eff$ was obtained as the ratio of the DPI and SPI cross sections derived in the leading collinear approximation of pQCD (LLA). 
This means that evolution of perturbative parton cascades was treated at the one-loop level, and the matrix elements of hard parton interactions treated in the Born approximation.  
In the DPI problem, the subleading non-logarithmic corrections to the LLA are bound to be sizable.   
Indeed, when deriving the total DPI cross section within the LLA accuracy, one integrates the differential distribution over jet imbalances, $\delta_{ik}^2 \ll Q_i^2$, up to the scale given by the transverse momenta of the jets, $Q_i^2$. 
In reality, due to experimental cuts that are imposed in order to extract jets in the {\em back-to-back}\/ kinematics the true hard scale of the DPI cross section is lower. 
Being formally a subleading $\cO{\as}$ correction, it will affect both the \12 and \22\ cross sections. Which way the subleading pQCD effects will change the ratio is so far unknown. 
To establish the true hard scales of the parton distributions entering the DPI cross section formula, one has to carry out the NLLA analysis which would include taking into consideration concrete details of the jet finding algorithms employed in the experimental setup. 

Finally, our prediction for the DPI cross sections was based on a model assumption of the absence of NP two-parton correlations in the proton. 
This assumption is arbitrary. One routinely makes it for the lack of any firsthand information about such correlations.  
In \cite{BDFS3} we have pointed out a source of genuine non-perturbative two-parton correlations that should come onto the stage 
for very small $x$ values, $x\ll 10^{-3}$, and estimated its magnitude via inelastic diffraction in the framework of the Regge--Gribov picture of high energy hadron interactions. 
Also it was argued in \cite{Schweitzer:2012hh} that strong quark--antiquark correlations may arise from dynamical chiral symmetry breaking.

\medskip

In order to be able to reliably extract the DPI physics, one has to learn how to theoretically predict  \oo\ parton collision processes with production of two hard systems (four jets in particular). 
This is the dominant channel, and it is only in the back-to-back kinematics that the \22\ and \12\ DPI processes become competitive with it. 
Among first subleading pQCD corrections to the \oo\ amplitude, there is a loop graph that looks 
like two-by-two parton collision. This resemblance is deceiving though. Unlike the \22\ and \12\ contributions, this specific correction does not depend on the spatial distribution of partons in the proton (information encoded in $\sigma_\eff$), is not power enhanced in the region of small transverse momenta of hard systems, and therefore does not belong to the DPI mechanism~\cite{BDFS2,Gaunt_1v2}.  
Treating the  \oo\ amplitude at the one-loop level corresponds to the two-loop accuracy for the cross section.  
Until this accuracy is achieved, the values of $\sigma_\eff$ extracted by experiments should be considered as tentative. 

\medskip
Our first conclusion is that in the kinematical region explored by the Tevatron experiments, the $x$-dependence of $\sigma_\eff$ turns out to be rather mild. 
This by no means implies, however, that $\sigma_\eff$ can be looked upon as any sort of a universal number. 
On the contrary, we see that the presence of the perturbative correlation due to the \12\ DPI mechanism results in dependence of  $\sigma_\eff$ not only on the parton momentum fractions $x_i$ and on the hardness parameters, but also on the type of the DPI process.

For example, in the case of golden DPI channel of production of two same sign $W$ bosons \cite{2W} the discussed mechanism leads to expectation of significantly larger $\sigma_\eff$ than for, say, $W$ plus two jets process.
Indeed, the comparison of the values of $R$ for central production of two gluon jet pairs, $Wjj$ and $W^+W^+$ (with jet transverse momenta $p_\perp\simeq M_W/2$), 
gives ($\sqrt{s}= 7\,\mbox{Te\!V} , \> \eta_1=\eta_2=0$)
\beq
\left.
\begin{array}{rl}
R(jj+jj)  & = 1.18 \> (0.81) \\
R(W+jj) & = 0.75 \> (0.45) \\
R(W^+W^+)  & = 0.49 \> (0.26)
\end{array}
\right\}
\>
 Q_0^2=0.5\>(1.0)\, \GeV^2.
\eeq
As a result of the varying magnitude of the perturbative correlation, the effective interaction areas $\sigma_\eff$ come out to be significantly different for the three processes:
\beq
\begin{array}{rl}
jj+jj : &\quad \sigma_\eff = 14.6 \div 17.6 \,\mbox{mb},\\
W+jj :&\quad \sigma_\eff = 18.3 \div 22.0 \,\mbox{mb},\\
W^+W^+ : &\quad \sigma_\eff = 21.5 \div 25.4 \,\mbox{mb}.
\end{array}
\eeq
The smaller value for each effective interaction area corresponds to more developed perturbative parton cascades 
than the larger one ($Q_0^2=0.5\,\GeV^2$ versus $Q_0^2=1.0\,\GeV^2$).

Contrary to the $W^+W^+$ channel, the {\em double Drell-Yan process}\/ favors the \12\ mechanism, $g\to u\bar{u}$. 
As a result, the effective interaction area here turns out to be even smaller. 

For central production of two $Z$ bosons at $\sqrt{s}= 7\,\mbox{Te\!V}$ we get
\beq\begin{split}
R(ZZ)  &= 1.03 \> (0.73), \cr
 ZZ : \quad \sigma_\eff &= 15.9 \div 18.5 \,\mbox{mb}. 
\end{split}
\eeq

An important feature of the \12 mechanism is its dependence on the hardness of the process. 
With increase of $Q_i^2$, the \12\ to \22\ ratio $R$ should increase rather fast thus pushing $\sigma_\eff$ to smaller values. 
At the same time, with decrease of the $p_\perp$ of the jets this contribution decreases.  
As we have seen above, such a trend is consistent with the D0 data for $x \sim 10^{-2}$.
 
By pushing the hardness scales down to $p_\perp \sim 3 \div 4\, \GeV$, one enters the domain of the physics of {\em minijets}. 
Here one should have $\sigma_\eff \sim $ 25  mb for $x_i\sim 10^{-2}$, the much larger value than 
 the $Q$-independent $\sigma_\eff$  which was assumed in the Monte Carlo models like PYTHIA for a long time.

It would be interesting to implement in the MC models a more realistic account of MPI in which $\sigma_{\eff}$ would decrease with increase of $p_\perp$.
\medskip

Acknowledgements:  This research was supported by the United States Department of Energy and the Binational Science Foundation.

\appendix
\section{Momentum sum rule \label{SEC:Momentum}}

The proof of the sum rule \eqref{eq:momsum} involves two types of graphs, see Fig.~\ref{momgraph}.

\begin{figurehere}
\centering
	 \includegraphics[width=0.45\textwidth]{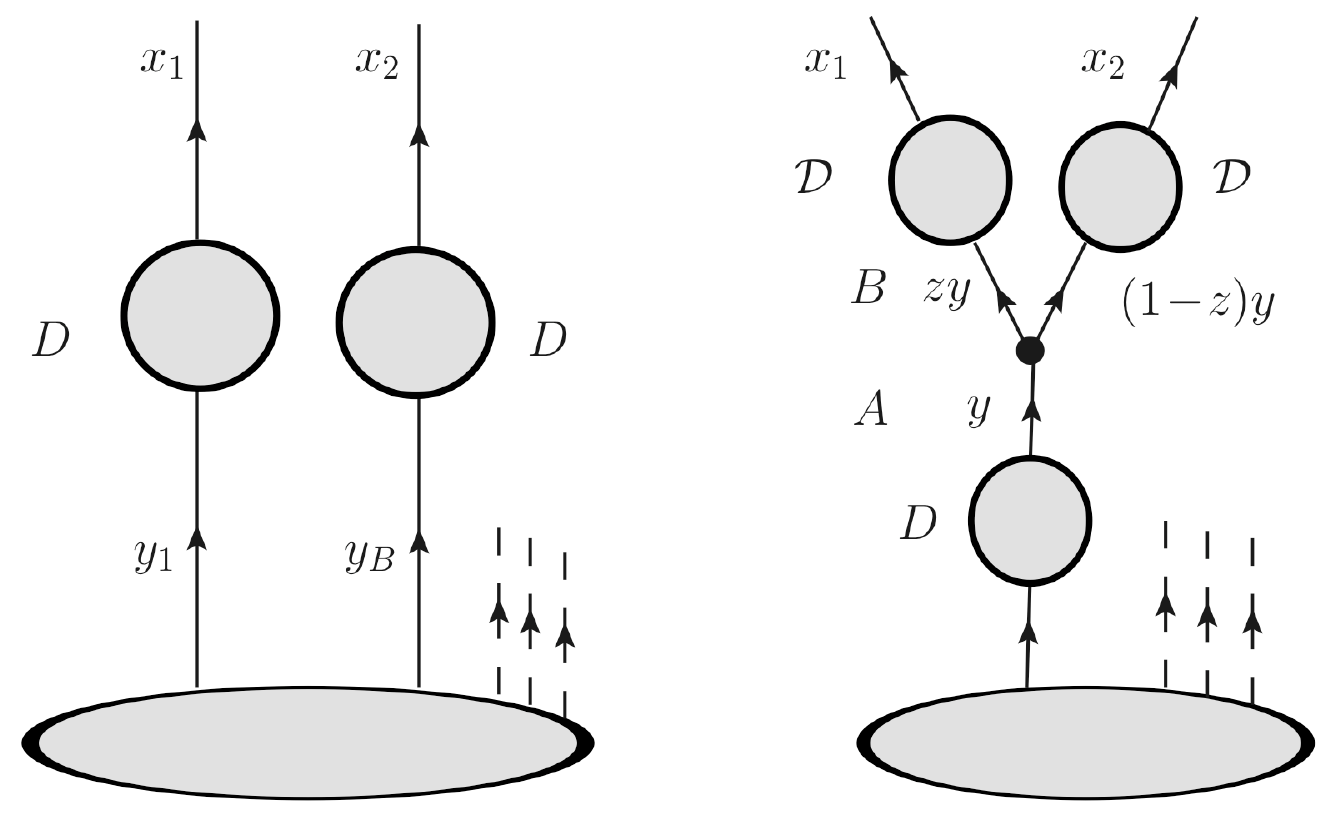}
\caption{\label{momgraph} 
Independent and correlation diagrams.}
\end{figurehere}

\subsection{Two partons from the wave function}

Consider first two partons 1 and 2 whose parents are taken from the hadron wave function at some scale $\Qsep2$ that separates NP and PT stages. 
Distribution of each parton then independently evolves up to the hard scale $Q^2$ according to the standard pQCD rules. 

Single parton distribution is presented in the form of convolution
\beq\nonumber
 D^f_h(x,Q^2) = \sum_B \int_x^1 \frac{dy}{y} \cD^f_i\left(\frac{x}{y}, Q^2;\Qsep2\right) w^i_h(y;\Qsep2)
\eeq
of the NP input function $w$ with the parton evolution function $\cD^f_i(x,Q^2;\Qsep2)$ obeying the initial condition
\beq\nonumber
\cD^f_i(x,\Qsep2;\Qsep2) = \delta^f_i \cdot \delta(1-x). 
\eeq
Applying the momentum integral \eqref{eq:momsum} to the single parton distribution $D^{i_2}(x_2,Q^2)$
gives 
\[
\sum_{i_2}   \int dx_2\, x_2 \, D^{i_2}_h(x_2) \>=\> \sum_B  \bar{y}_B, 
\]
with $\bar{y}_B$ the average energy fraction of the proton carried by the initial parton $B$ (parent of $i_2$) at the scale $\Qsep2$,
\[
 \bar{y}_B \>=\> \int dy_B \,y_B\,  w^B_h(y_B;\Qsep2). 
\]
Due to the momentum sum rule for the parton wave function of the proton, 
summing over parton species $B$ produces the total energy carried by {\em all initial partons}\/ but the parent of the second registered parton $i_1$:
\[
  \sum_B \bar{y}_B \>=\> 1- y_{1}.
\] 
The parent parton energy $y_1$ is not observable. What is fixed by the measurement is $x_1$, while $y_1$ is being integrated over:
\beq\label{eq:sumNP}
\begin{split}
& \sum_{i_2}  \int dx_2\, x_2 \, D^{i_1,i_2}_h(x_1,x_2) \cr
& = \sum_A \int_{x_1}^1\frac{dy_1}{y_1}  \bigg[ 1-y_1 \bigg] w_h^A(y_1) 
\cD_A^{i_1}\left(\frac{x_1}{y_1}, Q^2; Q_0^2 \right) \cr
 &\equiv \big[1- \left\langle y_1 \right\rangle \big] \cdot D_h^{i_1}\left(x_1, Q^2\right), 
\end{split}
\eeq
where we have introduced an average conditional parent parton energy $\left\langle y_1 \right\rangle$ which depends on $x_1$ and $Q^2$.
It is important to notice that this quantity depends also on the unphysical scale $\Qsep2$ which separates the domains of PT and NP description (unlike the physical inclusive parton distribution in the proton $D$ on the r.h.s.\ of \eqref{eq:sumNP}).

\subsection{One parton splitting into two}
For the perturbative correlation we have
\[
\begin{split}
\sum_{A,B} & \int\frac{dy}{y^2} \int\frac{dz}{z(1-z)} P_A^B(z)D^A_h(y)\cD_B^{i_1} \left(\frac{x_1}{zy} \right)\times (1-z)^2y^2 \cr
&= \sum_{A,B}  \int {dy} \int\frac{dz}{z} P_A^B(z)D^A_h(y)\cD_B^{i_1} \left(\frac{x_1}{zy} \right) \times (1-z).
\end{split}
\]
We split the factor $(1-z)$ into two pieces, $(1)+ (-z)$.
The first one gives
\beq\label{eq:difbel}
\begin{split}
 \frac{\as}{2\pi}   \sum_B & 
 \int {dy}  \int\frac{dz}{z} P_A^B(z)D^A_h(y)\cD_B^{i_1} \left(\frac{x_1}{zy} \right)  \cr
 = & -  \int  {dy}  S_A(k^2)  D^A_h(y, k^2) \cr
& \cdot \frac{\partial}{\partial \ln k^2} \biggl[ S_A^{-1}(k^2)\, \cD_A^{i_1} \bigl(\frac{x_1}{y}, Q^2; k^2 \bigr)\biggr].
\end{split}
\eeq 
Here we have used the evolution equation for the second D function differentiated over the {\em smaller}\/ scale $k^2$: 
\beq\label{eq:eveqB}
\begin{split}
& \frac{\as(k^2)}{2\pi} \sum_B  \int\frac{dz}{z} P_A^B(z)\, \cD_B^{p} \left(\frac{x}{z}, Q^2; k^2 \right)   \cr &
= \, -\> S_A(k^2)
\frac{\partial}{\partial \ln k^2} \bigg[ S_A^{-1}(k^2)\, \cD_A^{p} \big(x, Q^2; k^2 \big)\bigg] ,
\end{split}
\eeq
where $S_A$ is the Sudakov form factor of the parton $A$ depending on the two scales, the overall $Q^2$ and the floating splitting scale $k^2$ \cite{DokBook,DDT}. 

An alternative evolution equation where the derivative is applied to the {\em upper}\/ scale of the parton distribution in the proton reads
\beq\label{eq:eveqU}
\begin{split}
& \frac{\as(k^2)}{2\pi} \sum_A  \int\frac{dz}{z} P_A^B(z)\, D^A_{h} \left(\frac{x}{z}, k^2 \right)   \cr &
= \,  S_B^{-1}(k^2)
\frac{\partial}{\partial \ln k^2} \bigg[ S_B(k^2)\, D^B_{h} \big(x, k^2 \big)\bigg] .
\end{split}
\eeq
This equation allows us to analogously represent the second piece ($-z$) as the derivative of the first D-function over the upper scale:
\beq\label{eq:difupp}
\begin{split}
 \frac{\as}{2\pi} & \sum_A 
\int {dy} \,(-z) \int\frac{dz}{z} P_A^B(z)\,D_h^A(y) \cD_B^{i_1}(\frac{x_1}{zy}) \cr
 =&  -  \int {dy'} \, \cD_B^{i_1}\left(\frac{x_1}{y'}, Q^2; k^2\right)  S_B^{-1}(k^2) \cr
& \cdot \frac{\partial}{\partial \ln k^2} \bigg[ S_B(k^2)D_h^B(y', k^2)  \bigg], 
\end{split}
\eeq
with $y'\equiv zy$.
Combining \eqref{eq:difbel} and \eqref{eq:difupp}, we get a full logarithmic scale derivative of the product of the D-functions:
\[ 
-\sum_A\int dy \int\frac{dk^2}{k^2}\cdot \frac{d}{d\ln k^2} \biggl[ D_h^A(y, k^2) \cD_A^{i_1}\left(\frac{x_1}{y}, Q^2; k^2\right)\biggr] .
\] 
Now we integrate over the intermediate virtuality and make use of the boundary conditions, 
\[
\cD_A^{p}\left(\frac{x}{y}, Q^2; Q^2\right) = \delta_A^p\, \delta(1-\frac{x}{y}), \quad
D_h^A(y, \Qsep2) = w_h^A(y), 
\]
with $w_h^A$ the NP input parton distribution.  
We obtain
\beq\label{eq:pfin}
\begin{split}
& \sum_A \int_{x_1}^1 dy_1 \, w_h^A(y_1)\, \cD_A^{i_1}\bigl(\frac{x_1}{y_1}, Q^2; \Qsep2\bigr)  \cr
& -x_1 D_h^{i_1}(x_1, Q^2)  \> \equiv\> \big[ \left\langle y_1 \right\rangle -x_1\big] \cdot D_h^{i_1}(x_1, Q^2) . 
\end{split}
\eeq
Once the two- and one-parton contributions \eqref{eq:sumNP} and \eqref{eq:pfin} are taken together, 
the unphysical quantity $ \left\langle y_1 \right\rangle$ cancels out, and we arrive at the desired sum rule \eqref{eq:momsum}.

We conclude that for consistency of the DPI picture, the perturbative parton correlation (and thus the \12\  subprocesses) 
should be taken into full consideration at that very moment when one allows distributions of partons picked from the hadron wave function to evolve with the hard scale(s).

\end{document}